\documentclass[12pt,fleqn]{article}

\usepackage{amsmath}
\usepackage{amsthm}
\usepackage{amssymb}
\usepackage{amsfonts}
\usepackage{graphicx}
\usepackage{hyperref}
\usepackage{color}
\usepackage{tikz}

\newtheorem{lem}{Lemma}

\newtheorem{thm}{Theorem}

\newtheorem{cor}{Corollary}

\newtheorem{prop}{Proposition}

\theoremstyle{remark}

\newtheorem{remark}{Remark}
\theoremstyle{definition}

 \numberwithin{equation}{section}

\numberwithin{equation}{section}

\newcounter{comment}
\begin{document}

\title{Painlev\'{e} III asymptotics of Hankel determinants for a  perturbed Jacobi weight}

\author{Zhao-Yun Zeng$^a$, Shuai-Xia Xu$^b$\footnote{Corresponding author (Shuai-Xia Xu).
 {\it{E-mail address:}} {xushx3@mail.sysu.edu.cn} }, and Yu-Qiu Zhao$^a$}
  \date{ {\it{$^a$Department of Mathematics, Sun Yat-sen University, GuangZhou
510275, China}}\\
{\it{$^b$Institut Franco-Chinois de l'Energie Nucl\'{e}aire, Sun Yat-sen University, GuangZhou
510275,  China}}
}
\maketitle

\begin{abstract}

We study the Hankel determinants associated with the weight
$$w(x;t)=(1-x^2)^{\beta}(t^2-x^2)^\alpha h(x),~x\in(-1,1),$$
where   $\beta>-1$, $\alpha+\beta>-1$, $t>1$, $h(x)$ is analytic in a domain containing  $[-1,1]$ and  $h(x)>0$ for $x\in[-1,1]$.
In this paper,  based on the  Deift-Zhou nonlinear  steepest descent analysis, we study the double scaling limit of the Hankel determinants as
 $n\to \infty$ and $t\to 1$.  We obtain the asymptotic approximations of the Hankel determinants, evaluated in terms of the Jimbo-Miwa-Okamoto $\sigma$-function for the Painlev\'{e} III equation. The asymptotics of the leading coefficients and the recurrence coefficients for the perturbed Jacobi polynomials are also obtained.

\end{abstract}

%%%%%%%%%%%%%%%%%%%%%%%%%%%%%%%%%%%%%%%%%%%%%%%%%%%%%%%%%%%%%%%%%%

\vspace{5mm}



\section{Introduction and statement of results}

Let $w(x;t)$ be the perturbed Jacobi weight
\begin{equation}\label{pJacobi-weight}
w(x;t)=(1-x^2)^{\beta}(t^2-x^2)^\alpha h(x),~x\in(-1,1),
\end{equation}
where  $\beta>-1$, $\alpha+\beta>-1$, $t>1$, the function $h(x)$ is analytic in a domain containing  $[-1,1]$ and  $h(x)>0$ for $x\in[-1,1]$.
We study the Hankel determinants
\begin{equation}\label{Hankel}
D_n[w(x;t)]=\det(\mu_{j+k})_{j,k=0}^{n-1},
\end{equation}
where $\mu_i$ is the $i$-th moment of $w(x;t)$, namely,
 \begin{equation*}
 \mu_i=\int_{-1}^1 x^{i}w(x;t)dx,~~i=0,1,\cdots.
 \end{equation*}
The Hankel determinants possess the well-known multiple integral representation \cite[(2.2.11)]{Szego},
\begin{equation*}\label{Hankel representation}
D_n[w(x;t)]=\frac{1}{n!}\int_{[-1,1]^n}\prod_{i=1}^n w(x_i,t)\prod_{i<j}|x_i-x_j|^2\prod_{i=1}^n d x_i.
\end{equation*}
Via the above integral representation, the Hankel determinants are closely related to various fundamental quantities in random matrix
theory, such as the partition function, the gap probability of eigenvalues and the moment generating function of a certain random variable
associated with the random matrix ensemble; see \cite{mehta}. For example, in the Jacobi unitary ensemble corresponding to the weight
$w_1(x)=(1-x^2)^{\alpha}$, it is well-known that the probability distribution of the largest eigenvalues is
\begin{align*}\label{large eigenvalue}
  P_n(\lambda_{max}<s)&=\frac{1}{n! D_n[w_1(x)]}\int_{[-1,s]^n}\prod_{i=1}^n(1-x_i^2)^{\alpha}\prod_{i<j}|x_i-x_j|^2\prod_{i=1}^n d x_i \nonumber\\
&=\frac{\left(\frac{1+s}2\right)^{ n^2+2n\alpha}}{n!D_n[w_1(x)]}\int_{[-1,1]^n}\prod_{i=1}^n(1+x_i)^{\alpha}(\varpi(s)-x_i)^{\alpha}\prod_{i<j}|x_i-x_j|^2\prod_{i=1}^n d x_i,
\end{align*}
with $\varpi(s)=\frac{3-s}{1+s}$; see \cite{mehta}. Also, there is the remarkable Tracy-Widom formula for the
large-$n$ asymptotics of the distribution of the extreme eigenvalues  near the hard edge,
\begin{equation}\label{large eigenvalue asymptotic}
 \lim_{n\rightarrow\infty}-\frac{d}{d s}\ln P_n(\lambda_{max}<1-\frac{s}{2n^2})=\frac{\sigma_{\rm{JM}}(s)}{s};
\end{equation}see \cite{Tracy:Widom},
where $\sigma_{\rm{JM}}$  satisfies the Jimbo-Miwa-Okamoto $\sigma$-form of the Painlev\'{e} III equation  (\cite[(3.13)]{Jimbo})
\begin{equation}\label{PIII sigma function}
  (s\sigma_{\rm{JM}}'')^2+\sigma_{\rm{JM}}'(\sigma_{\rm{JM}}-s\sigma_{\rm{JM}}')(4\sigma_{\rm{JM}}'-1)-\alpha^2\sigma_{\rm{JM}}'^2=0
\end{equation}
and the boundary conditions
\begin{equation}\label{PIII sigma function boundary conditions}
 \sigma_{\rm{JM}}(s)\thicksim\frac{1}{4^{\alpha+1}\Gamma(1+\alpha)\Gamma(2+\alpha)}s^{1+\alpha},~s\rightarrow 0;~~
 \sigma_{\rm{JM}}(s)\thicksim\frac{s}{4}-\frac{\alpha}{2}\sqrt{s},~s\rightarrow\infty.
\end{equation}
It is worth  noting  that the Tracy-Widom formula \eqref{large eigenvalue asymptotic} holds for a large family of unitary ensembles, including the modified Jacobi unitary ensemble associated with the weight
\begin{equation*}
(1-x)^{\alpha}(1+x)^{\beta}h(x),\quad x\in(-1,1),~~\alpha>-1,~~\beta>-1.
\end{equation*}
The phenomenon is termed  universality in random matrix theory.

In \cite{for:wit}, Forrester and  Witte apply  the Okamoto  $\tau$-function theory to study the  random matrix average for the Laguerre unitary
ensemble
\begin{equation*}
   E(s,n)=\frac {1}{Z_n}\int_{[s,\infty)^n}\prod_{i=1}^n (x_i-s)^{\beta} x_i^{\alpha}e^{-x_i}\prod_{i<j}|x_i-x_j|^2\prod_{i=1}^n d x_i,
\end{equation*}
where $Z_n$ is the  normalization  constant.
It is shown that  the logarithmic derivative of the average $E(s,n)$ satisfies the Jimbo-Miwa-Okamoto $\sigma$-form of  the Painlev\'{e} V equation, with parameters depending on $n$.
By taking the scaling $s=\frac{ t}{4n}$ and letting $n\to \infty$, it is found that the  Painlev\'{e} V equation degenerates to a general  Jimbo-Miwa-Okamoto $\sigma$-form of the Painlev\'{e} III equation. Thus the  hard edge limiting average is obtained, generalizing the results of Tracy and Widom \eqref{large eigenvalue asymptotic}-\eqref{PIII sigma function boundary conditions}. The boundary conditions of the Painlev\'{e} III equation have also been studied in the    follow-up paper \cite{for:wit 2} of the same authors.

Now we mention several weights  closely related to \eqref{pJacobi-weight}.  A decade ago, in \cite{Kui:Mcl:Van :Vanl,Kui:Vanl}, Kuijlaars\;{\emph{et al.}\;considered  the orthogonal polynomials associated with the weight
\begin{equation*}
(1+x)^{\alpha}(1-x)^{\beta}h(x),\quad  x\in(-1,1),
\end{equation*}
which, in the case $\alpha=\beta$, is  the weight  \eqref{pJacobi-weight} with $t=1$.
The main focus of \cite{Kui:Mcl:Van :Vanl} is to obtain the asymptotics of the orthogonal polynomials, including those of the recurrence coefficients, the leading coefficients and the Hankel determinants.
In \cite{Kui:Vanl}, the results find applications  in random matrix theory, the Bessel limit kernel is obtained at the hard edge, and the kernel  is independent of the
perturbed analytic function $h$. Later, in \cite{Vanlessen2003},  Vanlessen studies the orthogonal polynomials associated with the further  generalized Jacobi-type weight with several singularities, of the form
 \begin{equation*}
 (1+x)^{\alpha}(1-x)^{\beta}h(x)\prod_{\nu=1}^p|x-x_{\nu}|^{2\lambda_{\nu}},\quad  x\in(-1,1),
 \end{equation*}
where $p$ is a fixed integer, $-1<x_1<x_2<\cdots<x_p<1$, $2\lambda_{\nu}>-1$, $\lambda_{\nu}\neq0$, $\alpha,~\beta>-1$,
and   $h$ is real analytic and strictly positive on $[-1,1]$. The asymptotics of the recurrence coefficients and the orthogonal polynomials associated are also obtained.

 Very recently, Basor, Chen and  Haq \cite{Basor:Chen:Ehrhardt} study the Hankel determinants associated with the weight
 \begin{equation*}
 \hat{w}(x,k)=(1-x^2)^{\beta}(1-k^2x^2)^{\alpha},\quad x\in(-1, 1), \quad \beta>-1,~\alpha \in \mathbb{R},
 \end{equation*}
 which is a special case of the weight \eqref{pJacobi-weight} with  $t=1/k$, and $h=k^{2\alpha}$.
For $n$ fixed, it is shown in  \cite{Basor:Chen:Ehrhardt},  via the ladder operator  method,    that the finite Hankel determinant  $D_n[\hat{w}(x;k)]$  is the $\tau$-function of the Painlev\'{e} VI equation; see also \cite{cz} for applications of the ladder operator method.  Large-$n$ asymptotics of the Hankel determinants are also obtained in \cite{Basor:Chen:Ehrhardt} for fixed $k$.

In this paper, however, we  focus on the asymptotics of the Hankel determinants, the leading coefficients,  and the  recurrence coefficients  associated with the weight \eqref{pJacobi-weight}, in the sense of a double scaling limit as $n\to \infty$ and $t\to 1$ when the algebraic singularity $x=t$  approaches the hard edge $x=1$.

Remarkable progress has been made in the study of the double scaling limit of  Hankel determinants and Toeplitz determinants
owning to the   Riemann-Hilbert approach developed by Deift, Zhou\;{\it et al.}\;\cite{deift,dkmv1,dkm}.
A series of questions and conjectures arose in the analysis of the Ising model (see \cite{Dei:Its:Kra survey}) and random matrices have been solved.    For example, in \cite{Cla:Its:Kra,Dei:Its:Kra}, Claeys, Deift and  co-authors obtain  Painlev\'{e} V asymptotics  of Toeplitz determinants and Hankel determinants  associated with  the  emergence of a Fisher-Hartwig singularity in the weight function. More recently, Claeys and Krasovsky \cite{Cla:Kra} study a weight with merging    Fisher-Hartwig     singularities and again a Painlev\'{e} V function is involved  to describe the transition between two different types of asymptotic behavior  of the Toeplitz determinants.

Other types of singularities have also been encountered.  In \cite{Bri:Mez:Mo},  Brightmore,  Mezzadri and Mo
consider the asymptotics of the partition function associated with the Gaussian weight perturbed by an essential singularity and they get  Painlev\'{e} III type asymptotics. In \cite{Xu:Dai:Zhao, Xu:Dai:Zhao 2}, Xu, Dai and Zhao also obtain Painlev\'{e} III type asymptotics of the Hankel determinants associated with the Laguerre weight with an essential singularity at the hard edge. In the double scaling limit of Hankel determinants,
the appearance of Painlev\'{e} functions is of particular interests; cf., e.g., \cite{CKV,Dai:kui,ik,xz2011}. The reader is referred to the comprehensive survey paper \cite{Dei:Its:Kra survey}
for the historic background and updated  results on the theory of Hankel determinants and Toeplitz determinants with applications in the Ising model.

In the preceding  papers \cite{xu:zhao:2014,xz2013b}, the authors have studied the transition asymptotics of the eigenvalue correlation kernel for the perturbed
Jacobi unitary ensemble defined by the perturbed Jacobi weight given in \eqref{pJacobi-weight}, varying from the Bessel kernel $\mathbb{J}_{\beta}$
to $\mathbb{J}_{\alpha+\beta}$ as the parameter $t$ varies in $(1,d]$ for a fixed $d>1$.  A  new class of universal  behavior at the edge of
the spectrum for the modified Jacobi ensemble is obtained and described in terms of the generalized Painlev\'{e} V equation, which in this case is equivalent to the Painlev\'{e} III equation after a M\"{o}bius transformation.

In the present paper, we  focus on the asymptotic approximations of the Hankel determinants, the leading coefficients, and the recurrence coefficients
of the polynomials orthogonal with respect  to the weight \eqref{pJacobi-weight}, in the sense of a double scaling limit as $n\to \infty$ and $t\to1$.
To simplify our discussion, we consider the even weight function \eqref{pJacobi-weight} by assuming $h$ is even, then we have the recurrence relation
\begin{equation}\label{recurrence relation}
z\pi_n(z)=\pi_{n+1}(z)+b_{n-1}^2\pi_{n-1}(z)
\end{equation}
for monic orthogonal polynomials with respect to the perturbed Jacobi weight.

\subsection{Statement of results}

To state the main results, we need the  Jimbo-Miwa-Okamoto $\sigma$-form of the Painlev\'{e} III equation
\begin{equation} \label{PIII Jimbo}
  (s\sigma_{{\rm JM}}'')^2+\sigma_{{\rm JM}}'(\sigma_{{\rm JM}}-s\sigma_{{\rm JM}}')(4\sigma_{{\rm JM}}'-1)-c_2\sigma_{{\rm JM}}'^2-c_1\sigma_{\rm{JM}}'-c_0=0;
  \end{equation}cf. \cite[(3.13)]{Jimbo},
  where $c_2=(\alpha+\beta)^2$, $c_1=-\frac 12 \beta(\alpha+\beta)$ and $c_0=\frac {\beta^2}{16}$.

Our first result is on the Painlev\'e III asymptotic approximations of the Hankel determinants, in terms of the Jimbo-Miwa-Okamoto $\sigma$-notation.
\begin{thm}\label{thm: asym Hankel}
Let  $\beta>-1$,  $\alpha+\beta>-1$,  $t>1$,  and $D_n(t)$ be   the
Hankel determinants given in \eqref{Hankel}  corresponding to the weight $w(x;t)$  in \eqref{pJacobi-weight}. As $n\rightarrow\infty$ and $t\rightarrow1$,
 we have the following asymptotic expansion
\begin{align}\label{asy:hankel}
\ln D_n(t)&=(n+\alpha+\beta)V_0-\alpha\ln h(t)-\beta\ln h(1)+\frac{1}{2}\sum_{k=1}^{\infty}k V_k^2+\left [(\alpha+\beta)^2-\frac{1}{4}\right ]\ln\frac{n}{4}\nonumber\\
          &\quad-\left [n^2+2n(\alpha+\beta)+1\right ]\ln2+\left [n+\alpha+\beta+\frac{1}{2}\right ]\ln2\pi+2\ln\frac{G(\frac{1}{2})}{G(\alpha+\beta+1)}\nonumber\\
          &\quad-\frac{\alpha^2}{2}\ln t+\frac 12(n\ln\varphi(t))^2-4\int_0^{4n\ln\varphi(t)}\frac{\sigma_{{\rm JM}}(\frac {s^2}{16})}{s}ds+o(1),
\end{align}
where $V_k=\frac{1}{2\pi}\int_0^{2\pi}e^{-ki\theta}\ln h(\cos\theta)d\theta$, $k=0,1,\cdots$, $G(z)$ is the Barnes $G$-function defined in \eqref{barnes' G function}, $\varphi(t)=t+\sqrt{t^2-1}$, the Jimbo-Miwa-Okamoto $\sigma$-function $\sigma_{\rm{JM}}$ is analytic on $(0,+\infty)$ and solves  the equation  \eqref{PIII Jimbo}
with the boundary conditions
 \begin{equation}\label{thm sigma TW 0}
 \sigma_{{\rm JM}}(s)=\frac{\beta}{4(\alpha+\beta)}s+C(\alpha,\beta)\left(\frac{s}{4}\right)^{1+\alpha+\beta}+O(s^{2+\alpha+\beta})+O(s^2)~~\mbox{as}~~s\rightarrow 0
\end{equation}
and
\begin{equation}\label{thm sigma TW infty}
 % \sigma_{{\rm JM}}(s)=\frac{1}{4}s-\frac{\alpha}{2}\sqrt{s}+\frac{\alpha^2}{4}+\frac{\alpha}{16}s^{-1/2}+O(s^{-1}),
 % ~{\rm as}~s\rightarrow\infty.
  \sigma_{{\rm JM}}(s)=\frac{1}{4}s-\frac{\alpha}{2}\sqrt{s}+\frac{1}{4}(\alpha^2+2\alpha\beta)-\frac{\alpha(\beta^2-\frac 14)}{4\sqrt{s}}+O(s^{-1})~~\mbox{as}~~s\rightarrow\infty,
\end{equation}
with
\begin{equation*}
C(\alpha,\beta)=
\frac{\alpha\Gamma(1-\alpha-\beta)\Gamma(\beta+1)}{(\alpha+\beta)\Gamma(1-\alpha)\Gamma(\alpha+\beta+2)\Gamma(\alpha+\beta+1)}.
\end{equation*}
\end{thm}\vskip .3cm

\remark{For $\alpha+\beta=0$, the condition \eqref{thm sigma TW 0} is simplified to
\begin{equation*}
\sigma_{{\rm JM}}(s)=\frac s4+O(s^2)~~\mbox{as}~~s\rightarrow0.
\end{equation*}
}

\remark{For $\beta=0$, the theorem is reduced to the celebrated  Tracy-Widom formula for the
large-$n$ asymptotic   distribution of the largest eigenvalues near the hard edge; cf.  \eqref{large eigenvalue asymptotic}-\eqref{PIII sigma function boundary conditions}, see also \cite{Tracy:Widom}.}\vskip .3cm

Our second result is on the transition asymptotics of the leading coefficients of the corresponding orthonormal polynomials, where the  Jimbo-Miwa-Okamoto $\sigma$-function is also involved.

\begin{thm} \label{thm:Asym:leading coefficent}
 Let  $n\rightarrow\infty$ and $t\rightarrow1$, then we have the asymptotic approximation  of the leading coefficient of the  orthonormal polynomial of degree $n$
 with respect to \eqref{pJacobi-weight}
\begin{equation}\label{asym:leading coeff}
\frac{\gamma_n}{2^n}=\frac{1}{\sqrt{\pi}D_t(\infty)}\left (1+2\sqrt{2}\left (\frac \alpha2+\frac q s\right )\sqrt{t-1}
+c_n\varphi(t)^{-2\alpha}+O(t-1) +O\left (\frac {s}{n^2}\right )\right ),
\end{equation}
where $\varphi(t)=t+\sqrt{t^2-1}$, $D_t(\infty)=2^{-(\alpha+\beta)}e^{-\frac 12 V_0}\varphi(t)^{\alpha}$, $V_0=\frac{1}{2\pi}\int_0^{2\pi}\ln h(\cos\theta)d\theta$, $q(s)=4\sigma_{{\rm JM}} (\frac {s^2}{16})-\frac{ s^2}{16}-(\alpha+\beta)^2+\frac 14$, $s=4n\ln{\varphi(t)}$, $c_n$ is independent of $t$ such that  $c_n=O\left (n^{-2}\right )$, and  the error term  $O\left(\frac {s}{n^2}\right )$ is uniform for $t\in(1,d]$.
\end{thm}

The third  result is on the transition asymptotics of the recurrence  coefficients of the corresponding monic orthogonal polynomials; see \eqref{recurrence relation}.

\begin{thm} \label{thm:Asym:recurrence coefficent}
 Let  $n\rightarrow\infty$ and $t\rightarrow1$, then we have the asymptotic approximation of the recurrence coefficients
\begin{equation}\label{asym:recurrence coeff}
b_{n-1}^2=\frac{1}{4}-8\left(\frac q s\right)'(t-1)+O\left (\frac{t-1}{n}\right )+O\left(\frac {\sqrt{t-1}}{n^2}\right )+O\left(\frac 1{n^3}\right ),
\end{equation}
where
$q(s)=4\sigma_{{\rm JM}} (\frac {s^2}{16})-\frac{ s^2}{16}-(\alpha+\beta)^2+\frac 14$, $s=4n\ln{\varphi(t)}$, $\varphi(t)=t+\sqrt{t^2-1}$, the derivative is taken with respect to $s$ and the error term $O\left(\frac 1{n^3}\right )$ is uniform for $t\in(1,d]$.
\end{thm}

\remark{The variable $s=4n\ln{\varphi(t)}\sim 4\sqrt 2 \;n \sqrt{t-1}\in(0,\infty)$ describes the gap between the hard edge and the algebraic singularity of the weight function (\ref{pJacobi-weight}).  Theorems \ref{thm: asym Hankel}, \ref{thm:Asym:leading coefficent} and  \ref{thm:Asym:recurrence coefficent} describe
the transition of asymptotics of the Hankel determinants, the  leading coefficients and the recurrence coefficients associated with  weights having  two different hard edge singularities: On the one side it is of the form $(1-x^2)^{\alpha+\beta}$ as $s\to 0^+$, on the other side with  $(1-x^2)^{\beta}$ as $s\to \infty$. For fixed $s$ taken in the transition region $(0,\infty)$, we obtain  the Painlev\'{e} type transition asymptotics in the double scaling limit. Moreover,  as  $s\to 0$ and $s\to \infty$, the limiting asymptotics of these quantities agree with the known asymptotics for the modified Jacobi weight  with fixed singularities.
    }\vskip .3cm

 Let  $s\to 0$ or $s\to \infty$, we obtain from Theorem \ref{thm:Asym:leading coefficent} the asymptotics of the leading coefficients corresponding to Jacobi polynomials of different orders. We state the results in the following corollary, which are in consistence with those obtained in \cite{Kui:Mcl:Van :Vanl}.

\begin{cor}\label{cor: transion leading coeff}
(i) Let  $n\rightarrow\infty$ and $t\to 1$  such that $s=4n\ln{\varphi(t)}\rightarrow0^+$, we have the asymptotic approximation  of the leading coefficients of the orthonormal polynomials
\begin{equation}\label{transion leading coeff s to 0}
\frac{\gamma_n}{2^n}=\frac{1}{\sqrt{\pi}D_1(\infty)}\left(1-\frac{4(\alpha+\beta)^2-1+O\left (s^2\right )}{8n}
+O\left (\frac{1}{n^2}\right )\right).
\end{equation}
(ii) Let $n\rightarrow\infty$ and $t\to 1$  such that $s\rightarrow\infty$, we have the asymptotic approximation of the leading coefficients
\begin{equation}\label{transion leading coeff s to infty}
\frac{\gamma_n}{2^n}=\frac{1}{\sqrt{\pi}D_t(\infty)}\left(1-\frac{4\beta^2-1}{8n}
+o\left (\frac{1}{n}\right )\right).
\end{equation}
\end{cor}

Similarly, as $s\to 0$ or $s\to \infty$, we extract  from Theorem  \ref{thm:Asym:recurrence coefficent}  the asymptotics of the recurrence coefficients corresponding to Jacobi polynomials with different parameters.   The results  are also consistent with those obtained in \cite{Kui:Mcl:Van :Vanl}.

\begin{cor}\label{cor: transion recurrence coeff}
(i) Let $n\rightarrow\infty$ and $t\to 1$  such that $s\rightarrow0^+$, we have the following asymptotic approximation of the recurrence coefficients
\begin{equation}\label{transion recurrence coeff s to 0}
b_{n-1}^2=\frac{1}{4}-\frac{4(\alpha+\beta)^2-1+O(s)}{16n^2}+O\left (n^{-3}\right ).
\end{equation}
(ii) Let $n\rightarrow\infty$ and $t\to 1$  such that $s\rightarrow\infty$ and $\frac {s^2} n \to 0$, we have the asymptotic of the recurrence coefficients
\begin{equation}\label{transion recurrence coeff s to infty}
b_{n-1}^2=\frac{1}{4}-\frac{4\beta^2-1}%+O(\frac{1}{s})}
{16n^2}+o\left (\frac {1}{n^2}\right ).
\end{equation}
\end{cor}\vskip .3cm

%Similarly, as $s\to 0$ and $s\to \infty$, we get  the asymptotic of the recurrence coefficients corresponding to Jacobi polynomials of different order. We state the result in the corollary as follows, which are also in consistence with that obtained in \cite{Kui:Mcl:Van :Vanl}.

%\begin{cor}\label{cor: transion recurrence coeff}
%(i)Let  $s\rightarrow0^+$ and $n\rightarrow\infty$, we have the asymptotic of the recurrence coefficients
%\begin{equation}\label{transion recurrence coeff s to 0}
%b_{n-1}^2=\frac{1}{4}-\frac{4(\alpha+\beta)^2-1+O(s)}{16n^2}+O(n^{-3}).
%\end{equation}
%(ii)Let $s\rightarrow\infty$,  we have the asymptotic of the recurrence coefficients
%\begin{equation}\label{transion recurrence coeff s to infty}
%b_{n-1}^2=\frac{1}{4}-\frac{4\beta^2-1+O(\frac{1}{s})}{16n^2}+O(n^{-3}).
%\end{equation}
%\end{cor}

To prove  the main results, first we derive several differential identities for the leading coefficients and the logarithmic derivative of the Hankel determinants, relating to the solution of the matrix Riemann-Hilbert (RH) formulation   for orthogonal polynomials.
Then we make use of the results obtained in a preceding  paper of the authors using
 the Deift-Zhou steepest descent method for the RH problems \cite{deift,dkmv1,dkm}.   The derivation  is   given in \cite{xu:zhao:2014} with full details, and is briefly reviewed in Section \ref{sec-RH-analysis} below.

The rest of the paper is organized as follows.
In Section \ref{sec-Model-PV-rhp}, we formulate  the model RH problem for the  Painlev\'{e} III equation, which
is introduced by the authors earlier in \cite{xu:zhao:2014}. An equivalent $\sigma$-form of Painlev\'{e} III equation  is then  derived. In Section \ref{sec-OPs-rhp},
we prove the differential identities for the Hankel determinants and the leading coefficients in terms of the RH problem for the  orthogonal polynomials associated with the weight \eqref{pJacobi-weight}. The identities  are the starting points of our analysis in later sections. In Section \ref{sec-RH-analysis}, we outline  the notations and  formulas resulted from  the RH analysis, obtained previously by the authors  in \cite{xu:zhao:2014}. The proofs of Theorems \ref{thm: asym Hankel}, \ref{thm:Asym:leading coefficent}, \ref{thm:Asym:recurrence coefficent}
 are provided  in the last section, Section \ref{sec-thm-proof}.

%%%%%%%%%%%%%%%%%%%%%%%%%%%%%%%%%%%%%%%%%%%%%%%%%%%%%%%%%%%%%%%%%%%%%%%%%%%%%%%%%%%%%%%%%

\section{ Painlev\'{e} III equation and $\sigma$-form of Painlev\'{e} III equation } \label{sec-Model-PV-rhp}
In \cite{xu:zhao:2014}, to construct the local parametrix in the nonlinear steepest descent analysis of the RH problems,
Xu and Zhao  introduce a modified version of the Painlev\'{e} V equation which is equivalent to the Painlev\'{e} III equation after a M\"{o}bius transformation.
\begin{prop}\label{Painleve III: Xu zhao} (Xu and Zhao \cite{xu:zhao:2014}) Assume that  $y(s)$  solves
 \begin{equation}\label{nonlinear diff order 2-introduction} \frac {d^2y}{ds^2}-\frac{2y}{y^2-1}\left(\frac {dy}{ds}\right)^2+\frac{1}{s} \frac {dy}{ds} +\frac {y(y^2+1)}{4(y^2-1)}+\frac {y}{2s}+\alpha\frac{y}{s}+\left (\beta-\frac 12\right )\frac{y^2+1}{2s}=0,\end{equation} where $\alpha$ and $\beta$ are constants.
The equation is converted  to a generalized Painlev\'{e} V equation by putting   $\omega=y^2$, so that
\begin{equation}\label{modified Painleve V-introduction}
 \frac {d^2\omega} {ds^2} - \left ( \frac 1 {\omega-1} +\frac 1 {2\omega} \right ) \left (\frac {d\omega} {ds} \right )^2+\frac 1 s\frac {d\omega} {ds}  +\frac {(2\alpha+1)\omega} s+\frac {\omega(\omega+1)}{2(\omega-1)}\pm \left (\beta-\frac 12\right )\frac{\sqrt{\omega}}{s}(\omega+1)=0, \end{equation}
which is reduced to the classical Painlev\'{e} V equation for $\beta=\frac 12$. Applying  the M\"{o}bius transformation  $v(s)=\frac{y(s)+1}{y(s)-1}$ turns  the equation \eqref{nonlinear diff order 2-introduction} into the Painlev\'{e} III equation
\begin{equation}\label{Painleve III cononic-introduction}
\frac {d^2v}{ds^2}-\frac{1}{v}\left(\frac {dv}{ds}\right)^2+\frac{1}{s} \frac {dv}{ds}  +\frac {1}{s}\left ( \frac {\alpha-\beta}2 v^2+\frac {\alpha-\beta+1}2\right ) -\frac {v^3} {16}+\frac{1}{16v}=0.
\end{equation}
Moreover, the equation \eqref{nonlinear diff order 2-introduction} is the
 compatibility condition for the Lax pair
\begin{equation}\label{Lax pair-1-introduction}
 \Psi_\lambda(\lambda, s)= \left (\frac{s\sigma_3} 2+\frac{A(s)}{\lambda-\frac 12} +\frac{B(s)}{\lambda+\frac 12}+\frac{(\beta-\frac 12)\sigma_1} {\lambda}
  \right )\Psi(\lambda, s),                                                  \end{equation}
\begin{equation}\label{Lax pair-2-introduction}
 \Psi_s(\lambda, s)=\left( \frac{\lambda\sigma_3} 2+u(s)\sigma_1  \right )
 \Psi(\lambda, s) ,                                                 \end{equation}
 where  $ \sigma_1=    \begin{pmatrix}
                  0 &1 \\
                    1 & 0
  \end{pmatrix}$ and
 $\sigma_3=\begin{pmatrix}
  1 &0 \\ 0 & -1
\end{pmatrix}$ are  the Pauli matrices,
\begin{equation}\label{coefficient A-B-introduction}
A(s)=\sigma_1B(s)\sigma_1,~~\mbox{and}~~B(s)=\left(
              \begin{array}{cc}
                b(s)-\frac{\alpha}2 & -(b(s)-\alpha)y(s) \\[0.2cm]
                b(s)/y(s) & -b(s)+\frac{\alpha}2 \\
              \end{array}
            \right) , \end{equation}
with $y(s)$ being a particular  solution of \eqref{nonlinear diff order 2-introduction},  while  $b(s)$ and $u(s)$ are determined  by the   equations
\begin{equation}\label{nonlinear equations 2-introduction}s\frac {dy}{ds} =-\frac{s y}{2}+\frac{b(y^2-1)^2}{y} -\alpha (y^2-1)y-(\beta-\frac 12)(y^2-1)  \end{equation}
and \begin{equation}\label{coefficient u-introduction}u(s)=\frac{ b(s)/y(s)-(b(s)-\alpha)y(s)}{s}+\frac{\beta-\frac 12} {s}. \end{equation}
\end{prop}
\qed

In view of  the symmetry  $\sigma_1\Psi(-\lambda)\sigma_1=\Psi(\lambda)$, new Lax pair of differential equations are obtained by applying the transformation
 \begin{equation}\label{psi-0 def}
  \Psi_0(\zeta,s)=
e^{-\frac{\pi i}{4}\sigma_3}\zeta^{\frac{1}{4}\sigma_3}\frac{I+i\sigma_2}{\sqrt{2}}
\Psi\left (\sqrt{\zeta}, s\right )e^{\frac{\pi i}{4}\sigma_3},~~\arg \zeta\in(-\pi, \pi).
\end{equation}
Then a model RH problem for $ \Psi_0(\zeta,s)$ in the $\zeta$-plane  is formulated,  and its unique  solvability is  proved   for $s>0$; see \cite {xu:zhao:2014}.
For later use, we recall the asymptotic behavior of the  model RH problem for $ \Psi_0(\zeta,s)$ as follows:
 \begin{equation}\label{psi-0 at infinity}\Psi_0(\zeta)=\zeta^{\frac{1}{4}\sigma_3}\frac{I-i\sigma_1}{\sqrt{2}}
    \left (I+\frac{\frac {\sigma(s)}{s}\sigma_3+iu(s)\sigma_1}{\sqrt{ \zeta}}+O\left (\frac 1{\zeta}\right )\right)
  e^{\frac {s\sqrt{\zeta}}2\sigma_3}\end{equation} as $\zeta\to \infty$  for $\arg\zeta\in (-\pi, \pi)$,
  where $s\in (0, \infty)$,  the function $u$ is defined in \eqref{coefficient u-introduction}  and $\sigma$ is defined as
  \begin{equation}\label{sigma-function}
 \sigma(s)=(b(s)-  \alpha/2)s-(su)^2;
  \end{equation}
also,
\begin{equation}\label{psi-0 at 1/4}
\Psi_0(\zeta)= E_0\left (I+\sum_{k=1}^{\infty}E_k(\zeta-  1/4)^k\right )(\zeta-  1/4)^{\frac 1 2\alpha\sigma_3}
\end{equation}as $\zeta\to   1 /4$ for $\arg(\zeta- 1 /4) \in (-\pi, \pi)$, where the coefficients can be determined by substituting  \eqref{psi-0 at 1/4} into (\ref{Lax pair-1-introduction}) and (\ref{psi-0 def}). For example,    the leading coefficient is
\begin{equation}\label{psi-0 at 1/4£º E-i}
E_0=\sqrt{\frac{b-\alpha}{2\alpha}}\; 2^{-\frac 1 2\sigma_3}\left(
      \begin{array}{cc}
      1+y(s)  &  \frac{ib(s)}{y(s)(b(s)-\alpha)}+i\\
      i(y(s)-1)   &\frac{b(s)}{y(s)(b(s)-\alpha)}-1\\
      \end{array}
    \right),
\end{equation}
and the $(1,1)$ entry of $E_1$ can be represented as
\begin{equation}\label{psi-0 at 1/4£º E-1 :11 entry}
  (E_1)_{11}=-\frac 1{\alpha}\left (\sigma-su+\alpha^2+\beta^2-\frac 14\right ).
\end{equation}
It is also noted that (see \cite[Sec.\;2.1]{xu:zhao:2014},  with $\Theta=-\alpha$)
\begin{equation}\label{sigma derivative} \sigma'=b-\frac {\alpha}{2},\quad (su)'=\frac 12\left (\frac by+y(b-\alpha)\right ).\end{equation}

%%%%%%%%%%%%%%%%%%%%%%%%%%%%%%%%%%%%%%%%%%%%%%%%%%%%%%%%%%%%%%%%%%%%%%%%%%%%%%%%%%%%%%%%%

\subsection{Jimbo-Miwa-Okamoto $\sigma$-form of  the Painlev\'{e} III equation} \label{sec-OPs-rhp}

     We derive the  Jimbo-Miwa-Okamoto $\sigma$-form of  the Painlev\'{e} III equation.

Let
\begin{equation}\label{q-definition}
 q(s)=\sigma(s)-us,
\end{equation}
where $\sigma(s)$ and $u(s)$ appear in the coefficients of the asymptotic behavior  of $\Psi_0$ at infinity, given in  \eqref{psi-0 at infinity}, then by the relation
\begin{equation}\label{sigma-u}
s\sigma'-\sigma=(su)^2;\end{equation}cf.  \eqref{sigma-function}  and \eqref{sigma derivative},
we have
\begin{equation}\label{derivative of q/s}
  \left(\frac qs\right )'=u^2-u',
\end{equation}
and
\begin{equation}\label{derivative of q}sq'=\sigma+(su)^2-s(su)'.\end{equation}
Then taking derivative on both sides of  (\ref{derivative of q}), using the definition of $q$ and the equation
\begin{equation*}
  \frac {d^2(su)}{ds^2}=2u\sigma'+\frac 14\left (su-\beta+\frac 12\right );
\end{equation*} see \cite[Sec.\;2.1]{xu:zhao:2014}  with $\Theta=-\alpha$ and $\gamma=\beta-\frac 1 2$, and in view of \eqref{sigma-u},
we obtain
\begin{equation}\label{u in terms of q}
  u=\frac {\frac \beta 4-\frac 18-q''}{2q'+\frac s 4 }.
\end{equation}
Substituting  (\ref{u in terms of q})  into (\ref{derivative of q/s}), we get the following third order equation
\begin{equation}\label{third order equation}
 \frac {q'''}{q'+\frac s8 }-\frac {(q''-\frac \beta4+\frac 18)(q''+\frac \beta4+\frac 18)}{2(q'+\frac s8)^2}-2\left (\frac q s\right )'=0.
\end{equation}

By the transformation
\begin{equation}\label{transformation}
q(s)=4\sigma_{{\rm JM}}(s^2/16)-s^2/16-c_2+\frac 14,\end{equation}
we get from (\ref{third order equation}) the new third order equation
\begin{equation} \label{new third order equation}
 III:= 2s^2\sigma_{{\rm JM}}'\sigma_{{\rm JM}}''' - s^2 {\sigma_{{\rm JM}}''}^2 + 2s \sigma_{{\rm JM}}' \sigma_{{\rm JM}}'' - 8s {\sigma_{{\rm JM}}'}^3 + (4\sigma_{{\rm JM}}+s-c_2) {\sigma_{{\rm JM}}'}^2 +c_0=0,
  \end{equation}
where $c_2=(\alpha+\beta)^2$, $c_0=\frac {\beta^2}{16}$.

Let $II$ denote the left-hand side of Jimbo-Miwa-Okamoto $\sigma$-form \eqref{PIII Jimbo} of the Painlev\'{e} III equation, namely,
\begin{equation} \label{first integral }
II(s):=  (s\sigma_{{\rm JM}}'')^2+\sigma_{{\rm JM}}'(\sigma_{{\rm JM}}-s\sigma_{{\rm JM}}')(4\sigma_{{\rm JM}}'-1)-c_2\sigma_{{\rm JM}}'^2-c_1\sigma_{\rm{JM}}'-c_0,
  \end{equation}
where $c_0$ and $c_2$ are defined in \eqref{new third order equation}, and  $c_1$ is a certain constant to be determined, then
\begin{equation} \label{II and third order equation}
\left( \frac {II}{\sigma_{{\rm JM}}'^2}\right)'\frac {\sigma_{{\rm JM}}'^3}{\sigma_{{\rm JM}}''}+II=III.
  \end{equation}
From the above equality  we see  that    the third order nonlinear  equation  \eqref{new third order equation}   is equivalent to
  the  Jimbo-Miwa-Okamoto $\sigma$-form \eqref{PIII Jimbo}  of the Painlev\'{e} III equation.
   Indeed, if  $\sigma_{\rm{JM}}$ satisfies the  $\sigma$-form \eqref{PIII Jimbo} for arbitrary coefficient $c_1$, in view of   \eqref{first integral } we have    $II(s)=c \sigma_{\rm{JM}}'$ for a constant $c$. Substituting it into  \eqref{II and third order equation} then yields  \eqref{new third order equation}. Conversely, if  $\sigma_{\rm{JM}}$ solves \eqref{new third order equation},  then \eqref{II and third order equation} is reduced to
\begin{equation*}
  \frac {II'}{II}= \frac {\sigma_{{\rm JM}}''}{\sigma_{{\rm JM}}'}.
\end{equation*}
Solving this equation   gives
\begin{equation*}
 II(s)=c\sigma_{{\rm JM}}',
\end{equation*}where $c$ is a constant.
Hence  $\sigma_{\rm{JM}}$ satisfies the Jimbo-Miwa-Okamoto $\sigma$-form \eqref{PIII Jimbo} of the Painlev\'{e} III equation.  The coefficient
$c_1=-\frac 12 \beta(\alpha+\beta)$, as can be determined by substituting the behavior of $\sigma_{{\rm JM}}$  at infinity into \eqref{PIII Jimbo}; see \eqref{sigma TW infty}.

We summarize the above derivation   as follows:
\begin{prop}\label{sigma form of PIII}
Let  \begin{equation*}
 q(s)=\sigma(s)-s u(s),
\end{equation*}
where $\sigma(s)$ and $u(s)$ appear in \eqref{psi-0 at infinity}  describing   the    asymptotic behavior  of $\Psi_0$ at infinity, then $q(s)$ satisfies the third order nonlinear differential  equation
\begin{equation*}
 \frac {q'''}{q'+\frac s8 }-\frac {(q''-\frac \beta4+\frac 18)(q''+\frac \beta4+\frac 18)}{2(q'+\frac s8)^2}-2\left (\frac q s\right )'=0.
\end{equation*}
By the transformation
\begin{equation*}
q(s)=4\sigma_{{\rm JM}}(s^2/16)-s^2/16-c_2+\frac 14,\end{equation*}
the above third order equation is turned into
  the Jimbo-Miwa-Okamoto $\sigma$-form of the Painlev\'{e} III equation
\begin{equation}\label{pro:sigma PIII }
  (s\sigma_{{\rm JM}}'')^2+\sigma_{{\rm JM}}'(\sigma_{{\rm JM}}-s\sigma_{{\rm JM}}')(4\sigma_{{\rm JM}}'-1)-c_2\sigma_{{\rm JM}}'^2-c_1\sigma_{\rm{JM}}'-c_0=0,
  \end{equation}
  where $c_2=(\alpha+\beta)^2$, $c_1=-\frac 12 \beta(\alpha+\beta)$ and $c_0=\frac {\beta^2}{16}$.\qed
\end{prop}\vskip .3cm

Noting that for $\beta=0$,  the equation  \eqref{pro:sigma PIII } is reduced to the special  Jimbo-Miwa-Okamoto $\sigma$-form \eqref{PIII sigma function} of  the Painlev\'{e} III equation, as appeared in \cite{Tracy:Widom}.

%%%%%%%%%%%%%%%%%%%%%%%%%%%%%%%%%%%%%%%%%%%%%%%%%%%%%%%%%%%%%

In \cite[Prop.\;2]{xu:zhao:2014},
it is proved that the RH problem for $\Psi_0(\zeta,s)$ has  a unique solution for $s\in(0,\infty)$. The nonlinear steepest descent analysis of the RH problem for
$\Psi_0(\zeta,s)$ is also carried out as $s\rightarrow0$ and $s\rightarrow\infty$. As a by-product,  the asymptotics of the specific Painlev\'{e}
function are then obtained. We collect the results in the proposition that follows, obtaining  directly from Proposition 3 in \cite{xu:zhao:2014}.

\begin{prop}\label{boundary condition}
The functions  $\sigma(s)$ and $u(s)$ are analytic in $s\in(0,\infty)$. For these and several other auxiliary functions,  we have the asymptotic behavior as $s\to \infty$:
\begin{equation}\label{boundary condition at infty}
\begin{split}
&y(s)=\pm\frac{2\beta-1}{s}+O\left (\frac{1}{s^2}\right ),\\
&\sigma(s)=-\frac{\alpha}{2}s-\left (\beta-\frac 12\right )^2-\frac{4\alpha(\beta-\frac 12)^2}{s}+O\left (\frac{1}{s^2}\right ),\\
&b(s)=\frac{4\alpha(\beta-\frac 12)^2}{s^2}+O\left (\frac{1}{s^3}\right ),\\
&u(s)=\frac{\beta-\frac 12}{s}+\frac{4\alpha(\beta-\frac 12)}{s^2}+\frac{16\alpha^2(\beta-\frac 12)}{s^3}+O\left (\frac{1}{s^4}\right ).
\end{split}
\end{equation}
As $s\to 0$, they behave as
\begin{equation}\label{boundary condition at 0}
\begin{split}
 & y(s)=1+\frac{\beta-\alpha}{2(\alpha+\beta)}s+O(s^2)+O(s^{2+2(\alpha+\beta)}),\\
& \sigma(s)=-(\alpha+\beta)^2-\frac{1}{4}+\frac{\alpha}{\alpha+\beta}\Big(-\frac{s^2}{32}+2C_0(\alpha,\beta)
  \Big(\frac{s^2}{16}\Big)^{1+\alpha+\beta}\Big)(1+O(s^2)),
\\
& b(s)=-\frac{(\alpha+\beta)^2}{s}+\frac{\alpha}{2}+\frac{3}{2}\frac{\alpha}{\alpha+\beta}\Big(-\frac{s}{16}
 +4C_0(\alpha,\beta)\Big(\frac{s^2}{16}\Big)^{\frac{1}{2}+\alpha+\beta}\Big)(1+O(s^2)),\\
& su(s)=-\frac{1}{2}-\frac{\alpha}{\alpha+\beta}\Big(-\frac{s^2}{32}+2C_0(\alpha,\beta)
  \Big(\frac{s^2}{16}\Big)^{1+\alpha+\beta}\Big)(1+O(s^2)),
\end{split}
\end{equation}
where
\begin{equation*}\label{coeff in sigma}
C_0(\alpha,\beta)=\frac{1}{2^{2+2(\alpha+\beta)}}
\frac{\Gamma(1-\alpha-\beta)\Gamma(\beta+1)}{\Gamma(1-\alpha)\Gamma(\alpha+\beta+2)\Gamma(\alpha+\beta+1)}.\hfill\qed
\end{equation*}
\end{prop}\vskip .3cm

As a corollary of Proposition \ref{boundary condition}, we have the following asymptotic behavior of  $\sigma_{\rm{JM}}$.

\begin{cor}\label{PIII JM boundary condition}
The Jimbo-Miwa-Okamoto $\sigma$-function $\sigma_{\rm{JM}}(s)$ is analytic for $s\in(0,\infty)$ and satisfies the boundary conditions
 \begin{equation}\label{sigma TW 0}
  \sigma_{{\rm JM}}(s)=\frac{\beta}{4(\alpha+\beta)}s+C(\alpha,\beta)\left(\frac{s}{4}\right)^{1+\alpha+\beta}+O(s^{2+\alpha+\beta})+O(s^2)~~
  \mbox{as}~s\rightarrow 0^+,
\end{equation}
and
\begin{equation}\label{sigma TW infty}
 % \sigma_{{\rm JM}}(s)=\frac{1}{4}s-\frac{\alpha}{2}\sqrt{s}+\frac{\alpha^2}{4}+\frac{\alpha}{16}s^{-1/2}+O(s^{-1}),
 % ~{\rm as}~s\rightarrow\infty.
  \sigma_{{\rm JM}}(s)=\frac{1}{4}s-\frac{\alpha}{2}\sqrt{s}+\frac{1}{4}(\alpha^2+2\alpha\beta)-\frac{\alpha(\beta^2-\frac 14)}{4\sqrt{s}}+O(s^{-1})
  ~~\mbox{as}~s\rightarrow +\infty,
\end{equation}
with
$$ C(\alpha,\beta)=
\frac{\alpha\Gamma(1-\alpha-\beta)\Gamma(\beta+1)}{(\alpha+\beta)\Gamma(1-\alpha)\Gamma(\alpha+\beta+2)\Gamma(\alpha+\beta+1)}.$$
\end{cor}

The asymptotic behavior of the  $\sigma$-function $\sigma_{\rm{JM}}(s)$ has been considered in \cite{for:wit 2} and \cite{Jimbo}. To compare the results with those obtained in \cite{Jimbo},  we take $\zeta(s)=-\sigma_{\rm{JM}}(4s)+s$, then by  \eqref{pro:sigma PIII } we have
\begin{equation}
  (s\zeta'')^2=4\zeta'(\zeta'-1)(\zeta-s\zeta')+\left ((\alpha+\beta)\zeta'-\alpha\right )^2.
  \end{equation}
The related $\tau$-function is defined  as
\begin{equation*}
 \zeta(s)=s\frac d{ds}\ln \tau(s)-\alpha\beta +s
\end{equation*}
in \cite{Jimbo} in our notations.
Then by \eqref{sigma TW 0}, we get the asymptotic of the  $\tau$-function
\begin{equation}\label{asy tau}
 \tau(s)=c s^{\alpha\beta} \left [1-\frac{\beta s}{\alpha+\beta}-\frac{C(\alpha,\beta)}{1+\alpha+\beta}s^{1+\alpha+\beta}+O(s^{2+\alpha+\beta})+O(s^2)
 +O(s^{2(1+\alpha+\beta)})\right ]
  \end{equation}
  with an arbitrary  constant $c$.
 The result is in consistence with \cite[Thm.\;3.2]{Jimbo}. It is noted that, in  \cite{Jimbo},  more general $\tau$-function  with three complex parameters is considered, thus  more restrictions on the parameters are needed.
  Yet in \cite{Jimbo}, in  our notation, a  restriction $-1<\mathrm{Re} (\alpha+\beta)\leq 0$ is brought in, and  the error estimate  therein is given as $O\left (s^{2(1+\mathrm{Re} (\alpha+\beta))}\right )$; see  \cite[Thm.\;3.2]{Jimbo}.

%%%%%%%%%%%%%%%%%%%%%%%%%%%%%%%%%%%%%%%%%%%%%%%%%%%%%%%%%%%%%%%%%%%%%%%%%%%%%%%%%%%%%%%%%

\section{Riemann-Hilbert  problem for orthogonal polynomials and differential identities} \label{sec-OPs-rhp}

Let  $\pi_n(z)$ be the monic polynomial of degree $n$ with respect to the weight $w(x)=w(x;t)$ in \eqref{pJacobi-weight},
then
\begin{equation}\label{Y-solution}
Y(z)= \left (\begin{array}{cc}
\pi_n(z)& \frac 1 {2\pi i}
\int_{-1} ^{1}\frac {\pi_n(s) w(s) }{s-z} ds\\[0.2cm]
-2\pi i \gamma_{n-1}^2 \;\pi_{n-1}(z)& -   \gamma_{n-1}^2\;
\int_{-1} ^{1}\frac {\pi_{n-1}(s) w(s) }{s-z} ds \end{array} \right ),
\end{equation}
is the unique matrix-valued function  analytic in   $\mathbb{C}\backslash [-1,1]$, fulfilling
  the jump condition
 \begin{equation}\label{Y-jump}
  Y_+(x)=Y_-(x) \left(
                               \begin{array}{cc}
                                 1 & w(x) \\
                                 0 & 1 \\
                                 \end{array}
                             \right)~~\mbox{for}~
      x\in (-1,1),\end{equation}
  the asymptotic condition at infinity
  \begin{equation}\label{Y-infty}
  Y(z)=\left (I+O\left (  1 /z\right )\right )\left(
                               \begin{array}{cc}
                                 z^n & 0 \\
                                 0 & z^{-n} \\
                               \end{array}
                             \right) \quad \mbox{as}\quad z\rightarrow
                             \infty ,\end{equation}
 and certain behavior demonstrating weak singularities  at $z=\pm 1$; see  \cite{fik} and \cite{xu:zhao:2014}.

To derive the asymptotic behavior  of the Hankel determinants, the leading coefficients and the recurrence coefficients,  we establish   differential identities to represent several   quantities  in terms of the matrix-valued function $Y$ in \eqref{Y-solution}.

\begin{lem}\label{lem: h-n relate Y} Let
\begin{equation} \label{h-n definition}
   h_n=\gamma_n^{-2}=\int_{-1}^1 \pi_n^2(x)w(x)dx,\end{equation}
 then $h_n$ can be expressed in terms of $Y(\pm t)$ as
\begin{equation} \label{h-n and Y}
  \frac d{dt}h_n=2\pi i \alpha(Y_{11}(-t)Y_{12}(-t)-Y_{11}(t)Y_{12}(t)),\end{equation}
 where $\gamma_n$ is the leading coefficient of the orthonomal polynomial of degree $n$, and $Y_{ij}(z)$ denotes the $(i,j)$ entry of $Y(z)$ given in \eqref{Y-solution}.
 \end{lem}

\noindent {\sc{Proof}}.
Taking derivative with respect to $t$ on both sides of \eqref{h-n definition} and making use of the orthogonality, we arrive at
 \begin{equation*}
 \frac d{dt}h_n=\int_{-1}^1 \pi_n^2(x)\frac \partial {\partial t}w(x)dx.
 \end{equation*}
It follows from \eqref{pJacobi-weight} that
 \begin{equation*}
 \frac {\partial}{\partial t}w(x;t)=\alpha w(x;t)\left (\frac {1}{t-x}+\frac{1}{t+x}\right ).
 \end{equation*}
Then the lemma is obtained by partial fraction decomposition of $\pi_n(x)/(x\pm t)$ and again  using the orthogonality.
\hfill\qed\\

For the Hankel determinants, we also have
\begin{lem}\label{lem: hakel:differential identities}
Let
\begin{equation} \label{H-n}
   H_n=\frac d{dt}\ln D_n(t), \end{equation}
then the following differential identity holds,
\begin{equation} \label{hakel:differential identities}
H_n=\alpha\left ( (Y^{-1}Y'_z)_{11}(t) -(Y^{-1}Y'_z)_{11}(-t)\right ).
\end{equation}
\end{lem}

\noindent {\sc{Proof}}.
By the well-known relation between the Hankel determinants and the leading coefficients
\begin{equation} \label{}
  \gamma_k^2=D_{k}/D_{k+1},~~D_n=\gamma_{n-1}^{-2}  \gamma_{n-2}^{-2}\cdots \gamma_{0}^{-2}; \end{equation}
 cf. Szeg\H{o} \cite[(2.2.15)]{Szego},  and the orthogonal relation
\begin{equation} \label{}
  \int_{-1}^1p_n(x)^2w(x)dx=1,\end{equation}
  where $p_n(z)=\gamma_n\pi_n(z)$ is the orthonormal polynomial with respect to \eqref{pJacobi-weight}, we get
\begin{equation} \label{Integal H-n}H_n=\int_{-1}^{1}\sum_{k=0}^{n-1}\gamma_k^2\pi_k^2(x) \frac {\partial w(x)}{\partial t}dx
=\alpha \int_{-1}^{1}\sum_{k=0}^{n-1}\gamma_k^2\pi_k^2(x)  w(x)  \left (\frac {1}{t-x}+\frac{1}{t+x}\right )  dx
.\end{equation}
Now the Christoffel-Darboux formula (\cite[(3.2.4)]{Szego}) applies and we have
\begin{equation} \label{CD formula}
 \sum_{k=1}^{n-1}p_k^2(x)= \gamma_{n-1}^2\left(\pi_{n-1} \frac d{dx} \pi_n-\pi_n \frac d{dx} \pi_{n-1} \right) \end{equation}
Substituting  \eqref{CD formula} into  \eqref{Integal H-n}   and using the  fraction decomposition techniques and   the orthogonal relation, we obtain \eqref{hakel:differential identities}.
\hfill\qed\\

%%%%%%%%%%%%%%%%%%%%%%%%%%%%%%%%%%%%%%%%%%%%%%%%%%%%%%%%%%%%%%%%%%%%%%%%%%%%%%%%%%%%%%%%%%%%%%%%%%%

\section{Nonlinear steepest descent analysis} \label{sec-RH-analysis}

The nonlinear steepest descent analysis  for the orthogonal polynomials has been provided    by two of the present authors in \cite[Sec.\;3]{xu:zhao:2014}.
The central piece  is the construction of  the local parametrix in a domain containing  singularities $z=1$ and $z=t$, in which  a modified
Painlev\'{e} V equation is involved. It is shown that the equation is equivalent to the Painlev\'{e} III equation after a M\"{o}bius transformation. In this section, we briefly review the results and collect  several formulas to be used in the investigation  of the Hankel determinants and the recurrence coefficients.

In \cite[Sec.\;3.4]{xu:zhao:2014}, applying a certain normalization at infinity, $Y(z)$ in \eqref{Y-solution} is approximated by
 \begin{equation}\label{global parametrix}
N_t(z)=   D_t(\infty)^{\sigma_3}M_1^{-1}a (z)^{-\sigma_3}M_1 D_t(z)^{-\sigma_3} \end{equation}
for $z$ kept away from   $[-1,1]$, where $M_1=\frac 1{\sqrt{2}}(I+i\sigma_1)$, $a
(z)=\left(\frac{z-1} {z+1}\right)^{1/4}$ for $z\in \mathbb{C}\backslash [-1, 1]$ with
branches   chosen such that $\arg (z\pm 1)\in (-\pi, \pi)$  and thus   $a (x) $ is  positive for $x>1$ and $a_+(x)/a_-(x)= i$ for $x\in (-1, 1)$, and the Szeg\H{o} function associated with  $w(x)$ takes the form
\begin{equation}\label{szego function}
D_t(z)=\left(\frac{z^2-1}{\varphi(z)^2}\right)^{\beta/2}\exp\left(\frac{\sqrt{z^2-1}}{2\pi}\int_{-1}^1\frac{\ln\left\{(t^2-x^2)^{\alpha}h(x)\right\}}{\sqrt{1-x^2}}\frac{dx}{z-x}\right), ~~z\in \mathbb{C}\backslash [-1, 1],
 \end{equation}in which $\varphi(z)=z+\sqrt{z^2-1}$ is analytic in $\mathbb{C}\setminus [-1,1]$ and $\varphi(z)\approx 2z$ as $z\to\infty$.
 %and $N_t$ solves the RH problem normalized at infinity, with jump on $(-1, 1)$
 %\begin{equation}\label{N-t jump} \left (N_t\right )_+(x)=\left (N_t\right )_-(x)\left(
  %     \begin{array}{cc}
   %    0 & w(x) \\
    %   - {w(x)}^{-1} & 0 \\
     %  \end{array}\right).
      %  \end{equation}
In \eqref{global parametrix},
\begin{equation}\label{szego function limit infty}
D_t(\infty)=\lim_{z\rightarrow\infty}D_t(z)=2^{-\beta}{\rm exp}\left(\frac{1}{2\pi}\int_{-1}^1\frac{\ln[(t^2-x^2)^\alpha h(x)] dx}{\sqrt{1-x^2}}\right).
\end{equation}

%\subsection{Local parametrix $P^{(1)}(z)$ at $z=t$ }

%The approximate solution $N_t$ fails to match $S$ at the end points $\pm1$.
%Thus local
%parametrices are to be constructed  in the neighborhood  $U(\pm1,\delta)$ containing the hard edge $\pm1$
%and the algebraic singularities $\pm t$ of the weight function.

%The RH problem for one of the local parametrix is as follows:
%\begin{itemize}
%  \item[(a)] $P^{(1)}(z)$ is analytic in $U(1,\delta) \backslash  \Sigma_{S}$;

 % \item[(b)] In  $U(1,\delta)$, $P^{(1)}(z)$ satisfies the same jump conditions as $S(z)$ does; cf. (\ref{S jump});

  %\item[(c)]  $P^{(1)}(z)$ fulfils the following  matching condition
  % on  $\partial U(1,\delta)$:
  %  \begin{equation}\label{matchingP1-N}
   %     P^{(1)}(z)N_t^{-1}(z)=I+ O\left ( n^{-1}\right ),~~\mbox{as}~n \to \infty.
    %\end{equation}
%\end{itemize}

In a disc   $U(1,\delta)$ centered at $z=1$ with fixed positive small radius $\delta$,  containing the hard edge $z=1$ and the algebraic singularity $ z=t$ of the weight function,
the local parametrix is constructed as
\begin{equation}\label{P-1-psi}
P^{(1)}(z)=E(z)\Psi_0\left (f_t(z),2n\sqrt{\rho_t }\right )\varphi(z)^{-n\sigma_3}W(z)^{-\frac 12\sigma_3} ,\end{equation}
where $W(z)=(z^2-1)^{\beta}(z^2-t^2)^{\alpha}h(z)$, $\arg(z\pm1)\in(-\pi,\pi)$, $\arg(z\pm t)\in(-\pi,\pi)$,
 $\Psi_0(\zeta)=\Psi_0(\zeta, s)$ is the solution to the model RH problem related to the Painlev\'{e} III equations; see \eqref{psi-0 def}.
Also, in \eqref{P-1-psi},
\begin{equation}\label{E}
 E(z)=N_t(z) W(z)^{\frac 12\sigma_3}\left\{G(f_t(z))\right\}^{-1},\end{equation}
where $G(\zeta)$ is a specific matrix function  defined as
\begin{equation}\label{G}G(\zeta)=\zeta^{\frac 14 \sigma_3}\frac {I-i\sigma_1}{\sqrt{2}}\exp\left\{ \left ( \frac {\alpha\sqrt\zeta} 2 \int^{\frac 1 4}_0 \frac 1 {\sqrt \tau } \frac {d\tau}{\tau-\zeta}\right )\sigma_3\right\},~~\zeta\in \mathbb{C}\backslash (-\infty, 1/4], \end{equation}
and the conformal mapping
\begin{equation}\label{conformal mapping}
f_t(z)=\frac{\left (\ln \varphi(z)\right )^2}{\rho_t}=\frac{2(z-1)}{\rho_t}(1+O(z-1)),\quad z\in U(1,\delta),
\end{equation}
 with  $\rho_t= 4 \left ( \ln \varphi(t)\right )^2$ such that $\rho_t=8(t-1)+O\left (  (t-1)^2\right )$ as $t\rightarrow 1$.

Accordingly,  $Y(z)$ in \eqref{Y-solution} is approximated  by the local parametrix in $U(1,\delta)$. More precisely, we have
\begin{equation} \label{Y in terms of psi}
   Y(z)=2^{-n\sigma_3}R(z)E(z)\Psi_0(f_t(z)) W(z)^{-\frac 12\sigma_3}, \quad t<z<1+\delta, \end{equation}
where $E$ and  $f_t$ is defined in \eqref{E} and \eqref{conformal mapping}, respectively,  and
\begin{equation}\label{R estimate}R(z)=I+O(n^{-1}),\end{equation} uniformly for $z$ in the whole complex plane.

%%%%%%%%%%%%%%%%%%%%%%%%%%%%%%%%%%%%%%%%%%%%%%%%%%%%%%%%%%%%%%

\section{Proof of the theorems} \label{sec-thm-proof}

In the present section, we apply the differential identities in Lemma \ref{lem: h-n relate Y} and Lemma \ref{lem: hakel:differential identities} and the results  obtained via the nonlinear steepest descent analysis for the orthogonal polynomials summarized  in Section \ref{sec-RH-analysis}, to derive the asymptotics of  the Hankel determinants, the leading coefficients and the recurrence coefficients.

%From the nonlinear steepest descent analysis,  by tracing back the invertible transformations \eqref{Y to T}, \eqref{T to S} and \eqref{S to R}, we have
%\begin{equation} \label{Y in terms of psi}
%   Y(z)=2^{-n\sigma_3}R(z)E(z)\Psi_0(f_t(z)) W(z)^{-\frac 12\sigma_3}, \quad t<z<\delta, \end{equation}
%where $E$ and  $f_t$ is defined in \eqref{E} and \eqref{conformal mapping}, respectively,  and $\Psi_0$ satisfies the model RH problem related to the Painlev\'{e} III equation   in Section \ref{sec-Model-PV-rhp}.

Substituting  \eqref{global parametrix} into  \eqref{E}, we have
\begin{equation}\label{E: iterms of NDG}
 E(z)=D_t^{\sigma_3}(\infty)M_1^{-1}a(z)^{-\sigma_3}M_1D_t(z)^{-\sigma_3}\left (W(z)\right )^{\frac 12\sigma_3}G^{-1}(f_t(z)),\end{equation}
 where $a(z)=\left (\frac {z-1}{z+1}\right )^{\frac 1 4}$ with $\arg(z\pm 1)\in (-\pi, \pi)$,  the Szeg\H{o} function $D_t(z)$,  and the auxiliary functions  $G$ and $f_t$ are defined in \eqref{szego function}, \eqref{G} and \eqref{conformal mapping}, respectively.

 Making a change of variables $s=\sqrt \tau$ in  the integral in \eqref{G}, we get a simpler representation for $G$, namely,
\begin{equation}\label{G: exp}
 G(\zeta)=\zeta^{\frac 14 \sigma_3}\frac {I-i\sigma_1}{\sqrt{2}}\left(\frac {2\sqrt{\zeta}-1}{2\sqrt{\zeta}+1}\right)^{\frac \alpha2\sigma_3}.
 \end{equation}

By the Cauchy theorem, the integral in \eqref{szego function} for the Szeg\H{o} function $D_t(z)$  can be written as the summation of the following two
integrals
\begin{equation}\label{Szego:Integral-1}
\begin{split}
\frac{\sqrt{z^2-1}}{2\pi}\int_{-1}^1\frac{\ln(t^2-x^2)^{\alpha}}{\sqrt{1-x^2}}\frac{dx}{z-x}=&\frac \alpha 2 \ln\frac{z^2-t^2}{\varphi(z)^2}-\frac {\sqrt{z^2-1}}2 \int_1^t\frac \alpha{\sqrt{x^2-1}}\frac{dx}{x-z} \\
& +\frac {\sqrt{z^2-1}}2\int_{-t}^{-1}\frac \alpha{\sqrt{x^2-1}}\frac{dx}{x-z}
\\
\end{split}
\end{equation}
and
\begin{equation}\label{Szego:Integral-2}
\frac{\sqrt{z^2-1}}{2\pi}\int_{-1}^1\frac{\ln h(x)}{\sqrt{1-x^2}}\frac{dx}{z-x}=\frac 12 \ln h(z)-\frac{\sqrt{z^2-1}}{4\pi i}\int_{\Gamma}\frac{\ln h(x)}{\sqrt{x^2-1}}\frac{dx}{x-z},
\end{equation}
where $\arg (z\pm 1)\in (-\pi,\pi)$ and $\Gamma$ is an anti-clockwise loop in the analytic domain of $h$ encircling   $[-1,1]$.
Then the integrals on the right-hand side of \eqref{Szego:Integral-1} are expressed explicitly in terms of elementary functions as
\begin{equation*}
\int_1^t\frac{1}{\sqrt{x^2-1}}\frac{dx}{x- z}=-\frac{1}{\sqrt{z^2-1}} \ln \left( \frac{zt+\sqrt{t^2-1}\sqrt{t^2-1}-1}{z-t} \right)
\end{equation*}
and
\begin{equation*}
\int_{-1}^{-t}\frac{1}{\sqrt{x^2-1}}\frac{dx}{x- z}=\frac{1}{\sqrt{z^2-1}}\ln \left( \frac{zt-\sqrt{t^2+1}\sqrt{t^2-1}-1}{z+t} \right),
\end{equation*}
where $\arg (z\pm1)\in(-\pi,\pi)$ and the  logarithmic functions take the principle branch.
Thus we get an explicit expression of the Szeg\H{o} function defined in \eqref{szego function}
 \begin{equation}\label{szego: function explicit}
 \begin{split}
 \frac{D_t^2(z)}{W(z)}=&\varphi(z)^{-2(\alpha+\beta)}\exp\left(-\frac{\sqrt{z^2-1}}{2\pi i}\int_{\Gamma}\frac{\ln
h(\zeta)}{\sqrt{\zeta^2-1}}\frac{d\zeta}{\zeta-z}\right)\\
&\times \left(\frac{z+t}{z-t}\; \frac{zt+\sqrt{z^2-1}\sqrt{t^2-1}-1}{zt-\sqrt{z^2-1}\sqrt{t^2-1}+1}\right)^{\alpha},
\end{split}
\end{equation}where in the fractional power we take the
  principle branches, and  $\arg (z\pm 1)\in (-\pi,\pi)$. It is readily verified that
$\left(\frac{D_t^2(x)}{W(x)}\right)_{+}\left(\frac{D_t^2(x)}{W(x)}\right)_{-}=1$ for $x\in (-1, 1)$.

In view of  \eqref{szego function limit infty} and using a  residue calculation argument,  we have
\begin{equation} \label{d-D(infty)}
 \frac d{dt}\ln D_t(\infty)=\frac{\alpha}{\sqrt{t^2-1}},\end{equation}
from which we obtain
\begin{equation} \label{D(infty)t and 1}
 D_t(\infty)=D_1(\infty)\left (t+\sqrt{t^2-1}\right )^{\alpha}.\end{equation}

Now a combination of  \eqref{E: iterms of NDG}, \eqref{G: exp} and  \eqref{szego: function explicit} gives
\begin{equation}\label{E-expand}
E(z)=D_t^{\sigma_3}(\infty)M_1^{-1}a(z)^{-\sigma_3}M_1\mathfrak{D}(z;t)^{\sigma_3}M_1\zeta^{-\frac{1}{4}\sigma_3},
\end{equation}
where
\begin{equation}\label{D hat}
\begin{split}
\mathfrak{D}(z;t) =& \varphi(z)^{\alpha+\beta}\exp\left (\frac{\sqrt{z^2-1}}{4\pi i}\int_{\Gamma}\frac{\ln
h(\zeta)}{\sqrt{\zeta^2-1}}\frac{d\zeta}{\zeta-z}\right )\\
& \times\left (\frac{(z+t)(\ln \varphi(z)-\ln \varphi(t))}{(z-t)(\ln \varphi(z)+\ln \varphi(t))}\right )^{-\frac{\alpha}{2}}
\left(
\frac{z t+\sqrt{z^2-1}\sqrt{t^2-1}-1}{z t-\sqrt{z^2-1}\sqrt{t^2-1}+1}\right)^{-\frac{\alpha}{2}},
\end{split}
\end{equation}
and $\zeta=f_t(z)$ is defined in \eqref{conformal mapping}.

Using Taylor expansions at $z=t$, we have
\begin{equation*}
 \left . \frac {\ln\varphi(z)-\ln \varphi(t)}{z-t}\right |_{z=t}=\frac{\varphi'(t)}{\varphi(t)}=\frac{1}{\sqrt{t^2-1}}.
\end{equation*}
Therefore,   at   $z=t$ we have
\begin{align}\nonumber
\mathfrak{D}(t;t)=\varphi(t)^{\alpha+\beta}\exp\left(\frac{\sqrt{t^2-1}}{4\pi i}\int_{\Gamma}\frac{\ln
h(\zeta)}{\sqrt{\zeta^2-1}}\frac{d\zeta}{\zeta-t}\right)
\left(\frac{\ln\varphi(t)}{t\sqrt{t^2-1}}\right)^{\frac{\alpha}{2}}.
\end{align}
By using the behaviors of $\varphi(t)$ and $\ln \varphi(t)$ as $t\to1$, and noting that
\begin{equation*}
  \frac{1}{\zeta-t}=\frac{1}{\zeta-1}\sum_{n=0}^{\infty}\left(\frac{t-1}{\zeta-1}\right)^n\quad \mbox{for}\quad
  \left|\frac{t-1}{\zeta-1}\right|<1,
\end{equation*}
we obtain
\begin{equation}\label{D hat t}
  \mathfrak{D}(t;t)=1+\sum_{k=1}^{\infty}d_k(t-1)^{\frac k2}
\end{equation}
for $ 0<t-1<\delta$,
where $\delta$ is a constant such that  $0<\delta<1$, the coefficients $d_k$ are explicitly computable and the first two  are
\begin{equation}\label{d-i}
d_1=\frac{\sqrt{2}}{2}(2(\alpha+\beta)+e_0)~~\mbox{and}~~ d_2=\frac{1}{4}(2(\alpha+\beta)+e_0)^2-\frac{2}{3}\alpha
\end{equation}
with
\begin{equation}\label{e-0}
  e_0=\frac 1{2\pi i}\int_{\Gamma}\frac{\ln
h(\zeta)}{\sqrt{\zeta^2-1}}\frac{ d\zeta}{\zeta-1}.
\end{equation}
For later use, we further compute the logarithmic derivative of $\mathfrak{D}(z;t)$. Repeatedly using Taylor expansions at $z=t$,
  we have
\begin{equation*}
\left . \frac d{dz}\ln\left(\frac {\ln\varphi(z)-\ln \varphi(t)}{z-t}\right)\right |_{z=t}
=\frac 1 2\left ( \frac {\varphi''(t)} {\varphi'(t)}- \frac {\varphi'(t)} {\varphi(t)}\right )
=-\frac t{2(t^2-1)}.
\end{equation*}
Similarly,   we get
\begin{equation*}
\frac d{dz}\ln\left(\frac {\ln\varphi(z)+\ln \varphi(t)}{z+t}\right)\Big|_{z=t}=\frac 1{2\sqrt{t^2-1}\ln \varphi(t)}-\frac 1{2t}=\frac 1{2(t^2-1)}-\frac 5{12}+O(t-1),
\end{equation*}
and
\begin{equation*}
\left. \frac d{dz}\ln\left ( z t+\sqrt{z^2-1}\sqrt{t^2-1}-1\right )\right|_{z=t}=\frac t{t^2-1}.
\end{equation*}
Putting these formulas together, taking logarithm of \eqref{D hat} and differentiating both sides,   and  in view of \eqref{D hat t},   we have
\begin{equation}\label{D-Hat: log derivative}
\left . \frac d{dz}\ln\mathfrak{D}(z;t)\right |_{z=t}=\left (\alpha+\beta+\frac{e_0}{2}\right )\frac 1{\sqrt{2(t-1)}}-\frac {\alpha} {3}+\sum_{k=1}^{\infty}d'_k(t-1)^{\frac k2},
\end{equation}
where $e_0$ is defined in \eqref{e-0} and $d'_k$ are  computable constants.

Substituting  \eqref{D hat t}  into \eqref{E-expand}, we get the expansion
\begin{equation}\label{E estimate}
  E(t)=D_t^{\sigma_3}(\infty)
  M_1\left(\sum_{k=0}^{\infty}C_{k}(t-1)^{k/2}\right) (2(t-1))^{\frac 14\sigma_3},
\end{equation}
where $M_1=\frac 1{\sqrt{2}}(I+i\sigma_1)$, the coefficients $C_k$ are explicitly computable matrices  and the first few are
 \begin{equation*}
 C_0=\left(
        \begin{array}{cc}
          1 & 0\\
          -\sqrt{2}id_1 & 1\\
        \end{array}
      \right) \quad\mbox{and}~~ C_1=\left(
                          \begin{array}{cc}
                            0 & 0 \\
                           \frac {2\sqrt{2}}{3} i\alpha & 0 \\
                          \end{array}
                        \right),
 \end{equation*}
 with $d_1$ defined in \eqref{d-i}.
By \eqref{psi-0 at 1/4£º E-i}, we get
\begin{equation}
E_0=\sqrt{\frac{b-\alpha}{2\alpha}}\sqrt{2}^{-\sigma_3}\left(
      \begin{array}{cc}
      1+y(s)  &  \frac{ib(s)}{y(s)(b(s)-\alpha)}+i\\
      i(y(s)-1)   &\frac{b(s)}{y(s)(b(s)-\alpha)}-1\\
      \end{array}
    \right),
\end{equation}
where $s=4n\ln \varphi(t)$.
From the boundary conditions   (\ref{boundary condition at infty}) and (\ref{boundary condition at 0}) for $y(s)$ and $b(s)$, we can derive the asymptotic behavior
\begin{equation} \label{error bound: at 0}
E_0=\sqrt{-(b-\alpha)}E_{0,0}(I +O(s)+O(s^{2(\alpha+\beta+1)}))  \quad as \quad s\to 0^+,
\end{equation}
and
\begin{equation} \label{error bound:at infty}
E_0=\sum_{k=0}^{\infty}E_{0,-k}s^{-k} \quad as \quad  s\to +\infty,
\end{equation}
where $E_{0,k}$ are computable constant matrices and
\begin{equation} \label{error bound:E-00}
(E_{0,0})_{21}=0.
\end{equation}

%%%%%%%%%%%%%%%%%%%%%%%%%%%%%%%%%%%%%%%%%%%%%%%%%%%%%%%%%%%%%%%%%%%%%%%%%%%%%%%%
\subsection{Proof of Theorem \ref{thm: asym Hankel}} \label{subsec-th3-proof}

Now we   substitute the asymptotics  we obtained for $Y$ into the differential identity \eqref{hakel:differential identities} for $\ln D_n(t)$. First, we consider the case for  $z$ close to $t$. From  \eqref{Y in terms of psi}, we obtain
\begin{equation*}
  \begin{split}
    Y^{-1}Y_z'=&W(z)^{\frac{1}{2}\sigma_3}\Psi_0^{-1}(\zeta)E^{-1}(z)E_z'(z)\Psi_0(\zeta)W(z)^{-\frac{1}{2}\sigma_3} +W(z)^{\frac{1}{2}\sigma_3}\Psi_0^{-1}(\zeta)\Psi_{0,z}'(\zeta)W(z)^{-\frac{1}{2}\sigma_3}\\
      &-\frac{W_z'}{2W}\sigma_3
      +W(z)^{\frac{1}{2}\sigma_3}\Psi_0^{-1}(\zeta)E^{-1}(z)R^{-1}(z)R'_z(z)
      E(z)\Psi_0(\zeta)W(z)^{-\frac{1}{2}\sigma_3},
  \end{split}
\end{equation*}
where $\zeta=f_t(z)$ is defined in \eqref{conformal mapping}.
Noting that $\zeta(t)=\left . \zeta\right |_{z=t}=\frac{1}{4}$. By using the behavior \eqref{psi-0 at 1/4} of $\Psi_0(\zeta)$ at $\zeta=1/4$,
and collecting \eqref{R estimate}, \eqref{E estimate} and \eqref{error bound: at 0}-\eqref{error bound:E-00} together, eventually we  have   the $(1,1)$ entry    of $Y^{-1}Y_z'$ at   $z=t$:
\begin{equation}\label{y:derivative:expand}
\begin{split}
  (Y^{-1}Y_z')_{11}(t)=&\left(\Psi_0^{-1}(\zeta(t))E^{-1}(t)E_z'(t)\Psi_0(\zeta(t))\right)_{11}
+\left .\left (\Psi_0^{-1}(\zeta)\Psi_{0,z}'(\zeta)\right )_{11}\right |_{z=t}\\
&-\left .\frac{1}{2}\frac{W_z'}{W}\right |_{z=t}+O\left (\frac {s^l}n\right )+O\left (\frac s {n}\right ),
\end{split}
\end{equation}
where %$\zeta=f_t(z)$ is defined in \eqref{conformal mapping},
$s=4n\ln \varphi(t)$, $l=\min(1,2(\alpha+\beta+1))$ and the error terms are uniform for $t\in(1,d]$.

Using \eqref{E-expand}, we further obtain
%It follows from \eqref{E-expand} that
\begin{equation}
  E^{-1}(t)E_z'(t)=\left(
                     \begin{array}{cc}
                     I_1(t) &I_2(t) \\
                       I_3(t) & -I_1(t) \\
                     \end{array}
                   \right),
\end{equation}
where
 \begin{align}
  &I_1(t)=\frac{a'(t)}{2a(t)}\left (\mathfrak{D}^2(t,t)+\mathfrak{D}^{-2}(t,t)\right )-\zeta_z'(t),
  \nonumber\\
   &I_2(t)=\frac{i}{2}\left (\frac{a'(t)}{2a(t)}\left (\mathfrak{D}^2(t,t)-\mathfrak{D}^{-2}(t,t)\right )+\left .\frac{d}{dz}\ln\mathfrak{D}(z,t)\right |_{z=t}\right ),
  \nonumber\\
  &I_3(t)=2i\left(\frac{a'(t)}{2a(t)}\left (\mathfrak{D}^2(t,t)-\mathfrak{D}^{-2}(t,t)\right )-\left . \frac{d}{dz}\ln\mathfrak{D}(z,t)\right |_{z=t}\right );\nonumber
  \end{align}
see \eqref{D hat} and \eqref{D hat t} for the definition of $\mathfrak{D}$.

It is readily seen that
\begin{equation}\label{derivative log a-t}
  \frac {d\ln a(t)} {dt}=\frac 1{2(t^2-1)},
\end{equation}and
\begin{equation}\label{derivative zeta}
 \left . \frac{d\zeta}{dz}\right |_{z=t}=\frac 1{2\sqrt{t^2-1}\ln\varphi(t)}.
\end{equation}
Then, from \eqref{D hat t}, \eqref{D-Hat: log derivative}, \eqref{derivative log a-t} and \eqref{derivative zeta}, we get the estimates as $t\to 1^+$, namely,
\begin{align}
   & I_1(t)=\frac{1}{4}\left ( \left (2(\alpha+\beta)+e_0\right )^2-\frac{1}{3}\right )+O(t-1),\label{estimate:I-1} \\
    &I_2(t)=i\left (-\frac{1}{3}\alpha+\frac{2(\alpha+\beta)+e_0}{2\sqrt{2}\sqrt{t-1}}\right )
+O\left (\sqrt{t-1}\right ),\label{estimate:I-2} \\
  &I_3(t)=O\left (\sqrt{t-1}\right ).\label {estimate:I-3}
\end{align}
From \eqref{psi-0 at 1/4}, we have
\begin{equation}\label{y:derivative:expand-term1}
 \Big(\Psi_0^{-1}(\zeta(t))E^{-1}(t)E_z'(t)\Psi_0(\zeta(t))\Big)_{11}=\left(E_0^{-1}\left(
                     \begin{array}{cc}
                     I_1(t) &I_2(t) \\
                       I_3(t) & -I_1(t) \\
                     \end{array}
                   \right)E_0\right)_{11},
\end{equation}
where $\zeta(t)=\frac 14$, and $E_0$ is defined in \eqref{psi-0 at 1/4£º E-i}.
Thus by \eqref{psi-0 at 1/4£º E-i} and \eqref{y:derivative:expand-term1}, we obtain
\begin{equation}\label{y:derivative:expand-term1:expand}\begin{aligned}
 \left (\Psi_0^{-1}(\zeta(t))E^{-1}(t)E_z'(t)\Psi_0(\zeta(t))\right )_{11}=&\frac{ s u-\beta+\frac{1}{2}}{\alpha}I_1(t)+\frac{2i(\sigma'(s)-(s u)')}{\alpha}I_2(t)\\
 & -\frac{i\left (\sigma'(s)+(s u)'\right )}{2\alpha} I_3(t),\end{aligned}
\end{equation}
where the derivative is taken with respect to $s$, $s=4n\ln \varphi(t)$.

By \eqref{psi-0 at 1/4}, we get
\begin{equation*}
  (\Psi_0^{-1}(\zeta)\Psi_{0,z}'(\zeta))_{11}=\zeta_z'({E}_1)_{11}+\frac{\frac{1}{2}\alpha\zeta_z'}{\zeta-\frac{1}{4}}.
\end{equation*}
Recalling    $W(z)=(z^2-1)^{\beta}(z^2-t^2)^{\alpha}h(z)$; cf. \eqref{P-1-psi},  we have
\begin{equation}\nonumber
  \frac{W_z'}{W}=\frac{2\beta z}{z^2-1}+\alpha\left(\frac{1}{z-t}+\frac{1}{z+t}\right)+\frac{1}{2}\frac{d}{dz}\ln h(z).
\end{equation}
Expanding the left-hand side  at $z=t$, we obtain
\begin{equation*}
  \left(\frac{\zeta_z'}{\zeta-\frac{1}{4}}-\frac{1}{z-t}\right )(t)=\frac{1}{2}\left(\frac{1}{\sqrt{t^2-1}\ln\varphi(t)}-\frac{t}{t^2-1}\right ).
\end{equation*}
Collecting these formulas together, and using \eqref{psi-0 at 1/4£º E-1 :11 entry}, we get
\begin{equation}\label{Y-11 Y-12 estimate second term}
 \begin{aligned} \left .\left((\Psi_0^{-1}(\zeta)\Psi_{0,z}'(\zeta))_{11}-\frac 12 \frac {W_z'}W\right)\right|_{z=t}=&
  -\frac{\sigma(s)-us+\beta^2-\frac{1}{4}+\frac {\alpha^2}2}{2\alpha\sqrt{t^2-1}\ln\varphi(t)} -\frac{\alpha}{4}\frac{t}{t^2-1}\\
& -\frac{\beta t}{t^2-1}-\frac{\alpha}{4t}-\left . \frac{1}{2}\frac{d}{dz}\ln h(z)\right |_{z=t}.
\end{aligned}
\end{equation}
%Here use has also been made of \eqref{psi-0 at 1/4£º E-1 :11 entry}.

Substituting (\ref{y:derivative:expand-term1:expand}) and (\ref{Y-11 Y-12 estimate second term}) into (\ref{y:derivative:expand}), we obtain
\begin{equation}\label{Y-11 Y-12 estimate at t}
 \begin{aligned}
 (Y^{-1}Y_z')_{11}(t)=&
  -\frac{   \sigma(s)-us+\beta^2-\frac{1}{4}+\frac {\alpha^2}2  }{2\alpha\sqrt{t^2-1}\ln\varphi(t)} -\left (\frac{\alpha}{4}+\beta\right )\frac{t}{t^2-1}
-\frac{\alpha}{4t}\\
& -\frac{1}{2}\frac{d}{dz}\ln h(z)\Big|_{z=t}+\frac{  s u-\beta+\frac{1}{2}  }{\alpha} I_1(t)+\frac{2i (\sigma'(s)-(s u)')   }{\alpha}I_2(t)\\
& -\frac{i(\sigma'(s)+(s u)') }{2\alpha}I_3(t)+O\left (\frac{s^l}{n}\right )+O\left (\frac{s}{n}\right ).
\end{aligned}
\end{equation}

Applying a similar argument to the case $z=-t$, we obtain
\begin{equation}\label{Y-11 Y-12 estimate at -t}
 \begin{aligned}
 (Y^{-1}Y_z')_{11}(-t)=&
  \frac{   \sigma(s)-us+\beta^2-\frac{1}{4}+\frac {\alpha^2}2      }{2\alpha\sqrt{t^2-1}\ln\varphi(t)} +\left (\frac{\alpha}{4}+\beta\right )\frac{t}{t^2-1}
+\frac{\alpha}{4t}\\
& -\frac{1}{2}\frac{d}{dz}\ln h(z)\Big|_{z=-t}+\frac{  s u-\beta+\frac{1}{2}  }{\alpha} \tilde{I}_1(-t)+\frac{2i (\sigma'(s)-(s u)')  }{\alpha}\tilde{I}_2(-t)\\
& -\frac{i (\sigma'(s)+(s u)') }{2\alpha}\tilde{I}_3(-t)+O\left (\frac{s^l}{n}\right )+O\left (\frac{s}{n}\right )
\end{aligned}
\end{equation}
with
 \begin{align}
& \tilde{I}_1(-t)=-\frac{1}{4}\left((2(\alpha+\beta)+e_0)^2-\frac{1}{3}\right )+O(t-1),\label{estimate:titlde I-1}\\
&\tilde{I}_2(-t)=-i\left (-\frac{1}{3}\alpha+\frac{2(\alpha+\beta )+e_0}{2\sqrt{2}\sqrt{t-1}}\right)
+O\left (\sqrt{t-1}\right ), \label{estimate:titlde I-2}\\
&\tilde{I}_3(-t)=O\left (\sqrt{t-1}\right ).
\label{estimate:titlde I-3}
\end{align}
Here use has been made of the fact that  $h(x)$ is an even function.

 Substituting (\ref{Y-11 Y-12 estimate at t}) and (\ref{Y-11 Y-12 estimate at -t}) into (\ref{hakel:differential identities})  yields
  \begin{equation}\label{hankel: log derivative proof}
 \begin{aligned}
 \frac{d}{d t}\ln D_n(t)=&\frac{1}{2}\left [(2(\alpha+\beta)+e_0)^2-\frac{1}{3}\right ](su-\beta+\frac{1}{2})-\frac{  \sigma(s)-us+\beta^2-\frac{1}{4}+\frac {\alpha^2}2 }{\sqrt{t^2-1}\ln\varphi(t)} \\
&-\left [\frac{2(\alpha+\beta)+e_0}{\sqrt{2}\sqrt{t-1}}
-\frac{2\alpha}{3}\right ]\left( \sigma'(s)-(s u)'\right ) -\frac{(\alpha^2+4\alpha\beta )t}{2(t^2-1)}  \\
&-\frac{\alpha^2}{2t}-\frac{\alpha}{2}\left (\frac{d}{dz}\ln h(z)\Big|_{z=t}-\frac{d}{dz}\ln h(z)\Big|_{z=-t}\right )
 \\
&+O\left (\frac{s^l}{n}\right )+O\left (\frac{s}{n}\right )+O\left(\sqrt{t-1}\right ),
\end{aligned}
\end{equation}
where $()'=\frac{d}{d s}$.
Integrating both sides of  this identity from $1+\varepsilon$ to some $t>1$ gives
\begin{equation}\label{hankel: log integral proof}
 \begin{aligned}
 \ln D_n(t)=&\ln D_n(1+\varepsilon)-\int_{4n\ln\varphi(1+\varepsilon)}^{4n\ln\varphi(t)}\frac{\sigma(s)-us+\beta^2-\frac{1}{4}+\frac {\alpha^2}2}{s}ds
 \\
&   -\left (\frac{1}{2}\alpha^2+2\alpha\beta\right )\int_{1+\varepsilon}^t\frac{t}{t^2-1}dt    -\frac{1}{2}\alpha^2\ln t+\frac{1}{2}\alpha^2\ln(1+\varepsilon) \\
& -\frac{\alpha}{2}V(t)-\frac{\alpha}{2}V(-t)+\frac{\alpha}{2}V(1+\varepsilon)+\frac{\alpha}{2}V(-1-\varepsilon)+R_n(t)+o(1),
\end{aligned}
\end{equation}holding uniformly for
 arbitrary  $ \varepsilon>0$,
where $V(z)=\ln h(z)$, and the remainder term
\begin{align*}
 R_n(t)&=\frac{1}{2}\left ((2(\alpha+\beta)+e_0)^2-\frac{1}{3}\right )\int_{1+\varepsilon}^t(su-\beta+\frac{1}{2})dt\nonumber\\
&\quad -\int_{1+\varepsilon}^t\left (\frac{2(\alpha+\beta)+e_0}{\sqrt{2}\sqrt{t-1}}
-\frac{2}{3}\alpha\right )\left (\sigma'(s)-(s u)'\right )dt.
 \end{align*}
Since $s=4n\ln\varphi(t)$, then for small $t-1$, we may approximate the integral
\begin{equation}\label{integral estimate in hankel}
 \int_1^t\frac{t}{t^2-1}dt%=\int_0^s\frac{t}{4n\sqrt{t^2-1}}ds
 =\int_0^s\frac{1}{s}ds+O(t-1).
\end{equation}
From (\ref{boundary condition at infty}) and  (\ref{boundary condition at 0}), namely  the boundary conditions for $u$ and $\sigma$,  we get the estimates for the integrals
\begin{align*}
   &\int_1^t \left (su-\beta+\frac{1}{2}\right ) dt%=\frac{1}{16n^2}\int_0^s(s u-\beta+\frac{1}{2})s ds
   =O(t-1),  \\
  & \int_1^t(\sigma-su)'dt%=\frac{1}{16n^2}\int_0^s s d(\sigma-su)
  =o(1),\\
  &\int_1^t\frac{1}{\sqrt{t^2-1}}(\sigma-su)'dt%=\frac{1}{4n}\int_0^s (\sigma-su)' d s
  =o(1),
\end{align*}
which give the estimate for $R_n(t)$ as $\varepsilon\rightarrow0^+$
\begin{equation}\label{Reminder estimate}
  |R_n(t)|=O(\sqrt{t-1})=o(1).
\end{equation}
Letting $\varepsilon\rightarrow0^+$, substituting  (\ref{integral estimate in hankel}) and (\ref{Reminder estimate}) into (\ref{hankel: log integral proof})
 and making use of the fact that $h(x)$ is an even function, we obtain
\begin{align}
 \ln D_n(t)&=\ln D_n(1)-\alpha V(t)+\alpha V(1)-\frac{1}{2}\alpha^2\ln t\nonumber\\
&\quad  -\int_0^{s}\frac{\sigma(s)-us+(\alpha+\beta)^2-\frac{1}{4}}{s}ds+o(1),
\label{hankel: log integral proof:2}
\end{align}
 where $s=4n\ln \varphi(t)$, $\varphi(t)=t+\sqrt{t^2-1}$.
 The convergence of the integral is guaranteed by the initial condition of $\sigma(s)$ and $u(s)$ in (\ref{boundary condition at 0}).

 The asymptotic approximation for $\ln D_n(1)$ has been  given in \cite[Thm.\;1.20]{Dei:Its:Kra} as
  \begin{align}
 \ln D_n(1)=&-\left (n^2+2n(\alpha+\beta)+1\right )\ln2+\left (\alpha+\beta)^2-\frac{1}{4}\right )\ln\frac{n}{4}\nonumber\\
 &+\left (n+\alpha+\beta+\frac{1}{2}\right )\ln2\pi
 +2\ln\frac{G(\frac{1}{2})}{G(\alpha+\beta+1)}\label{asym hankel t=1}\\
 &+(n+\alpha+\beta)V_0-(\alpha+\beta)V(1)
 +\frac{1}{2}\sum_{k=1}^{\infty}k V_k^2,\nonumber
\end{align}
where $V_k=\frac{1}{2\pi}\int_0^{2\pi}e^{-ki\theta}\ln h(\cos(\theta))d\theta$, $k=0,1,\cdots$. And the Barnes $G$-function is defined by the product
\begin{equation}\label{barnes' G function}
  G(1+z)=(2\pi)^{\frac z2}e^{-\frac {z+z^2(1+\gamma_E)}{2}}\prod_{k=1}^{\infty}\left(\left (1+\frac zk\right )^ke^{\frac {z^2}{2k}-z}\right),
\end{equation}
where $\gamma_E$ is the Euler constant. The  Barnes $G$-function satisfies the well-known recurrence relation
$$G(z+1)=G(z)\Gamma(z),$$
where $G(1)=1$, and $\Gamma(z)$ is the gamma function.

Substituting (\ref{asym hankel t=1}) into (\ref{hankel: log integral proof:2}) yields (\ref{asy:hankel}). Thus   completing  the proof of Theorem \ref{thm: asym Hankel}.

%%%%%%%%%%%%%%%%%%%%%%%%%%%%%%%%%%%%%%%%%%%%%%%%%%%%%%%%%%%%%%%%%%%%%%%%%%%%%%%%%%%%%%%%%%%
\subsection{Proof of Theorem \ref{thm:Asym:leading coefficent}} \label{subsec-th2-proof}

From  \eqref{psi-0 at 1/4},    \eqref{Y in terms of psi} and \eqref{R estimate}, we obtain
\begin{equation} \label{Y at t}
   Y(t)= 2^{-n\sigma_3}(I+O(  1/n))E(t)E_0\{l(t)\}^{\sigma_3},
\end{equation}
where $l(t)=\lim_{z\to t}(f_t(z)-\frac 14)^{\frac \alpha 2}W(z)^{-\frac 12}$, $E_0$  and $E(t)$ are defined in \eqref{psi-0 at 1/4£º E-i} and  \eqref{E estimate}, respectively.
Thus,  from \eqref{Y at t} and the differential identity \eqref{h-n and Y}, we have
\begin{equation}
\begin{aligned} \label{derivative h-n }
  \frac d {dt }h_n=&-\frac {\pi i\alpha}{2^{2(n-1)}}D_{t}(\infty)^2\Big (C_{12}^2(E_0)_{21}(E_0)_{22}\frac {1}{\sqrt{2(t-1)}}  \\
  &  +C_{12}C_{11}\big((E_0)_{12}(E_0)_{21}+(E_0)_{11}(E_0)_{22}\big)+C_{11}^2(E_0)_{11}(E_0)_{12}\sqrt{2(t-1)}\Big ),
\end{aligned}\end{equation}
where $C_{i,j}$ stands for the $(i,j)$ entry of the matrix
\begin{equation}\label{C}C=\left (I+O\left (\frac 1n\right )\right )M_1\sum_{k=0}^{\infty}C_k(t-1)^{\frac k2};
\end{equation}
see \eqref{E estimate} for the   constant matrices $M_1$, $C_0$ and $C_1$.
In view of  \eqref{psi-0 at 1/4£º E-i}, \eqref{sigma derivative}  and Proposition  \ref{Painleve III: Xu zhao},
we may write
\begin{equation}\label{E-ij product-1}
(E_0)_{21}(E_0)_{22}=\frac {2i}{\alpha}(\sigma-su)',\quad  (E_0)_{11}(E_0)_{12}=\frac {i}{2\alpha}(\sigma+su)',\end {equation}
and
\begin{equation}\label{E-ij product-2}
(E_0)_{12}(E_0)_{21}=\frac {1}{2\alpha}\left (su-\beta+\frac 12-\alpha\right ),\quad  (E_0)_{11}(E_0)_{22}=\frac {1}{2\alpha}\left (su-\beta+\frac 12+\alpha\right ).\end {equation}

Now, substituting \eqref{C}, \eqref{E-ij product-1} and \eqref{E-ij product-2} into \eqref{derivative h-n },
we obtain
\begin{align} \label{asymptotic d- gamma}
   \frac {d h_n}{dt }=&-\frac {2\sqrt 2\pi D_{t}(\infty)^2}{2^{2n}}\left\{\frac {(\sigma-su)'}{\sqrt{t-1}}\left (1+O\left (\frac 1n\right )+O(t-1)\right )            \right . \nonumber   \\
  & +(\sigma+su)'O(\sqrt{t-1}\; )\\
  &\left. -\left (su-\beta+\frac 12\right )\left (\frac 1{\sqrt{2}}+d_1+O\left (\frac 1n\right )+O(\sqrt{t-1}\; )\right )\right \},\nonumber
\end{align}
where $d_1$ is defined in \eqref{d-i} and
the error term $O(\frac 1 n)$ is uniform for $t\in(1, d]$.

%Thus \begin{equation} \label{asymptotic d- gamma-2}\frac d {dt }h_n=-\frac {2\sqrt 2\pi }{2^{2n}}D_{1}(\infty)^2(t+\sqrt{t^2-1})^{2\alpha}(\sigma'-(us)')\frac 1{\sqrt{t-1} }+O(1).\end{equation}

From  \eqref{boundary condition at infty} and \eqref{boundary condition at 0}, we can derive  the estimates of the integrals
\begin{equation}\label{error: integral 1}
 \int_{1}^t(su-\beta+\frac 12 )\sqrt{t-1}dt=O\left( \frac{t-1}{n}\right)+O\left((t-1)^{\frac 32}\right)
\end{equation}and
\begin{equation}\label{error: integral 2}
  \int_{1}^t(\sigma+su )'\sqrt{t-1}dt=o\left( \frac{t-1}{n}\right)+O\left((t-1)^{\frac 32}\right).
\end{equation}

Now, in view of \eqref{D(infty)t and 1},  \eqref{boundary condition at infty} and \eqref{boundary condition at 0},  and  integrating  by parts once, we have
\begin{align}\label{error: integral 3}
 \int_{1}^t\frac{D_{t}(\infty)^2(\sigma-su )'dt}{\sqrt{t-1}}=&2D_{t}(\infty)^2\left (\frac {\sigma}{s}-u\right )\left (\sqrt{t-1}+O\left ((t-1)^{\frac 32}\right )\right )\nonumber\\
 &-\frac {\sqrt{2}\alpha}{8n^2}\int_0^sD_{t}(\infty)^2(\sigma-su)ds+ O\left (\frac {t-1}{n}\right)\\
 &-\frac {\sqrt{2}D_{1}(\infty)^2}{4n}\left (\frac 14-(\alpha+\beta)^2\right )+O\left((t-1)^{\frac 32}\right).\nonumber
\end{align}
Then,   integrating   both sides of \eqref{asymptotic d- gamma} and using the estimates  \eqref{error: integral 1}-\eqref{error: integral 3}, we obtain
\begin{equation}
\begin{aligned} \label{asy of h-n}
h_n(t)=&h_n(1)-\frac {\pi D_{1}(\infty)^2}{2^{2n}} \frac {4(\alpha+\beta)^2-1}{4n}-\frac {\pi D_{t}(\infty)^2}{2^{2n}}\left\{  4\sqrt{2}\left(\frac {\sigma}{s}-u\right )\sqrt{t-1}\right .
 \\
& \left . +I(s)\frac1 {n^2} +O\left (\frac {\sqrt{t-1}}{n}\right )+O\left ((t-1)^{\frac 32}\right )\right\}, \end{aligned}\end{equation}
where
\begin{equation}\label{I-s}
 I(s)=-\frac{\sqrt{2}\alpha}{8D_{t}(\infty)^2} \int_{0}^s D_{t}(\infty)^2\left\{ (\sigma-su)+\frac {1+\sqrt{2}d_1}{4\alpha}\left (su-\beta+\frac 12\right )s\right\}ds.
 \end{equation}
 By   \eqref{boundary condition at infty} and \eqref{boundary condition at 0}, we get the estimate
\begin{equation}\label{I-s-estimate}
 I(s)
 =O(s^2)+O(s).
 \end{equation}
To determined $h_n(1)$, we use a result from \cite[Thm.\;1.6]{Kui:Mcl:Van :Vanl}, that is,
\begin{equation} \label{asymptotic:h-n at 1}
  h_n(1)=\frac {\pi}{2^{2n}}D_{1}(\infty)^2\left ( 1+ \frac {4(\alpha+\beta)^2-1}{4n} +c_n\right ),\quad c_n=O\left (\frac 1{n^2}\right ).\end{equation}
Substituting \eqref{I-s-estimate} and \eqref{asymptotic:h-n at 1} into \eqref{asy of h-n},  and noting that the relation between $D_1(\infty)$ and $D_t(\infty)$ \eqref{D(infty)t and 1}, we have
\begin{equation}\begin{aligned} \label{asy of h-n-formula}
h_n(t)=&\frac {\pi}{2^{2n}}D_{t}(\infty)^2\left \{1- \left [4\sqrt{2}\left(\frac {\sigma}s-u\right )+2\alpha\sqrt{2}\right ] \sqrt{t-1}+O\left ((t-1)^{\frac 32}\right )\right .\\
& \left . +O\left (\frac{s}{n^2}\right )+c_n \big (t+\sqrt{t^2-1}\; \big )^{-2\alpha}\right\}, \end{aligned}\end{equation}
where $c_n$ is independent of $t$ and  $c_n=O(\frac 1{n^2})$.
It follows from Proposition \ref{sigma form of PIII} that
\begin{equation}
 \sigma-su= 4\sigma_{{\rm JM}}\left(\frac {s^2}{16}\right )-\frac {s^2}{16}-(\alpha+\beta)^2+\frac 14.
 \end{equation}
Since $h_n=\gamma_n^{-2}$, we obtain the asymptotic approximation of the leading coefficient as in \eqref{asym:leading coeff}. Thus  completing the proof of Theorem \ref{thm:Asym:leading coefficent}.
%%%%%%%%%%%%%%%%%%%%%%%%%%%%%%%%%%%%%%%%%%%%%%%%%%%%%%%%%%%%%%
\subsection{Proof of Theorem \ref{thm:Asym:recurrence coefficent}} \label{subsec-th3-proof}

To approximate the recurrence coefficients, we recall their  relation with the leading coefficients
\begin{equation} \label{gamma and b-n}
  b_{n-1}^2=\frac { \gamma^2_{n-1}}{ \gamma_n^2}.\end{equation}

Let
\begin{equation}\label{lamda:1:hat:def}
\Lambda_1=\frac {\alpha} 2s+\sigma -us,
\end{equation}
then $\Lambda_1$ is analytic for $s\in(0,\infty)$ and
\begin{equation}\label{lamda:1:hat:initial}
 \Lambda_1(s)=\frac{1}{4}-(\alpha+\beta)^2+O(s^l) \quad \mbox{as} \quad s\to 0^+
\end{equation}with $l=\min(1,2(1+\alpha+\beta))$,
and
\begin{equation}\label{lamda:1:hat:infinty}
 \Lambda_1(s)=\frac{1}{4}-\beta^2+\sum_{k=1}^{\infty}\frac {l_{k}}{s^{k}} \quad \mbox{as} \quad s\to +\infty,
\end{equation}where $l_k$ are constant coefficients.
Then it follows from \eqref{lamda:1:hat:def}, \eqref{lamda:1:hat:initial} and \eqref{lamda:1:hat:infinty} that
\begin{equation} \label{lamda:hat:deff}
\Lambda_1(s-  s/n)=\Lambda_1(s)-\frac sn \Lambda_1'(s)+O\left (\frac {s^2}{n^2}\right ),\end{equation}
where  the error term is uniform for $s\geq0$.
From \eqref{lamda:hat:deff}, and using the fact that $s\thicksim4\sqrt{2}n\sqrt{t-1}$ as $t\to 1$, we have
\begin{align}\label{expansion of q/s}
  \frac{\Lambda_1(s-\frac sn)}{s-\frac sn}%&=[\Lambda_1(s)-\Lambda_1'(s)\frac{s}{n}+O(\frac{s^2}{n^2})]\frac{1}{s}[1+\frac{1}{n}+O(\frac{1}{n^2})]\nonumber\\
  =\frac{\Lambda_1}{s}-4\sqrt{2}\left(\frac{\Lambda_1}{s}\right)'\sqrt{t-1}+O
  \left(\frac{s}{n^2}\right).
\end{align}
Using the expression of $I(s)$ in \eqref{I-s}, and in view of \eqref{boundary condition at infty} and \eqref{boundary condition at 0}, we get
\begin{equation}\label{expansion of I-s}
\frac{ I(s-\frac{s}{n})}{(n-1)^2}=\frac{  I(s)}{n^2}+O\left(\frac{t-1}{n}\right).
\end{equation}
Then, a combination of  \eqref{asy of h-n}, \eqref{asymptotic:h-n at 1}, \eqref{expansion of q/s} and \eqref{expansion of I-s}  gives the following asymptotic formula for $h_{n-1}$
\begin{align} \label{asy of h-n-1-formula}
h_{n-1}(t)=&\frac {4\pi D_{t}(\infty)^2}{2^{2n}}\left\{1-4\sqrt{2}\frac{\Lambda_1(s)}{s} \sqrt{t-1}
+32\left(\frac{\Lambda_1}{s}\right)'(t-1)+\frac{I(s)}{n^2}\right .\nonumber\\
&\left .+c_n\left (t+\sqrt{t^2-1}\right )^{-2\alpha}+O\left(\frac{t-1}{n}\right)+O\left ((t-1)^{\frac 32}\right )
+O\left(\frac{\sqrt{t-1}}{n^2}\right)\right\}.
\end{align}
%It follows from   \eqref{boundary condition at infty}, \eqref{boundary condition at 0} and \eqref{I-s} that
%\begin{equation} \label{I-s:deff}
%\frac {I(s- \frac s n)}{(n-1)^2}=\frac {I(s)}{n^2}+O\left (\frac {t-1}{n}\right ).\end{equation}

From \eqref{asy of h-n}, \eqref{asy of h-n-1-formula} and \eqref{gamma and b-n}, %  \eqref{lamda:hat:deff} and \eqref{I-s:deff},
we obtain  the asymptotic approximation  of the recurrence coefficients stated in \eqref{asym:recurrence coeff}.
%%%%%%%%%%%%%%%%%%%%%%%%%%%%%%%%%%%%%%%%%%%%%%%%%%%%%%%%%%%%%%%%%%%%%%%%%%%%%%%%%%%%%%%%%%%%%%%

\section*{Acknowledgements}
The authors are grateful to the referees for their valuable suggestions and comments.
The work of Shuai-Xia Xu  was supported in part by the National
Natural Science Foundation of China under grant number
11201493, GuangDong Natural Science Foundation under grant number S2012040007824, and the Fundamental Research Funds for the Central Universities under grant number 13lgpy41.
 Yu-Qiu Zhao  was supported in part by the National
Natural Science Foundation of China under grant number
10871212.

%%%%%%%%%%%%%%%%%%%%%%%%%%%%%%%%%%%%%%%%%%%%%%%%%%%%%%%%%%%%%%%%%%%%%%%%%%%%%%%%%%%%%%%%%%

\end{document}